\documentclass[twocolumn]{aastex63}

\usepackage{natbib}
\usepackage{amsmath}
\usepackage{braket}
\usepackage{comment}
\usepackage{bm}
\usepackage{color}
\hypersetup{
 colorlinks=true,%
 linkcolor=black,
 citecolor=blue,
}

\def\shimizu#1{\textcolor{black}{{#1}}}

\received{}
\revised{}
\accepted{\today}
\submitjournal{ApJ}

\graphicspath{{./}{figure/}}

\begin{document}

\title{Role of Longitudinal Waves in Alfv\'en-wave-driven Solar Wind
}
\author{Kimihiko Shimizu}
\affiliation{School of Arts \& Science, The University of Tokyo, 3-8-1 Komaba, Meguro, Tokyo, 153-8902, Japan}
\author{Munehito Shoda}
\affiliation{National Astronomical Observatory of Japan, National Institutes of Natural Sciences, 2-21-1 Osawa, Mitaka, Tokyo, 181-8588, Japan}
\author{Takeru K. Suzuki}
\affiliation{School of Arts \& Science, The University of Tokyo, 3-8-1 Komaba, Meguro, Tokyo, 153-8902, Japan}
\affiliation{Department of Astronomy, The University of Tokyo, 7-3-1, Hongo, Bunkyo, Tokyo, 113-0033, Japan}

\begin{abstract}
    We revisit the role of longitudinal waves in driving the solar wind.
    We study how the the $p$-mode-like vertical oscillation on the photosphere affects the properties of solar winds under the framework of Alfv\'en-wave-driven winds. 
    We perform a series of one-dimensional magnetohydrodynamical numerical simulations from the photosphere to \shimizu{beyond several tens} of solar radii. 
    We find that the mass-loss rate drastically increases with the longitudinal wave amplitude at the photosphere up to $\sim 4$ times, in contrast to the classical understanding that the acoustic wave hardly affects the energetics of the solar wind.
    \shimizu{The addition of the longitudinal fluctuation induces the longitudinal-to-transverse wave mode conversion in the chromosphere,
    which results in the enhanced Alfv\'enic Poynting flux in the corona. Consequently, the coronal heating is promoted to give higher coronal density by the chromospheric evaporation, leading to the increased mass-loss rate.}
    This study clearly shows the importance of the longnitudinal oscillation in the photosphere and the mode conversion in the chromosphere in determining the basic properties of the wind from solar-like stars.
\end{abstract}


\section{Introduction} \label{sec:intro}
Inspired by the observed double-tail structure of comets, which indicates the presence of gas outflow \citep{Biremann_1951ZA}, \citet{Parker_1958ApJ} predicted the outward expansion of the hot coronal plasma, which results in the formation of transonic outflow. 
Later on, the in-situ measurement by the Mariner 2 Venus probe confirmed the existence of supersonic plasma streams from the Sun, which is now called the solar wind \citep{Neugebauer_1966JGR}. 
Hot coronae and stellar winds are also ubiquitously observed in low-mass main sequence stars that possess a surface convection zone \citep{Wood_2005ApJ,Wood_2021ApJ,Gudel_2014_proceedings,Vidotto2021LRSP}

In the framework of the thermally-driven wind model, the energy source of the solar wind is the thermal energy of the solar corona.
The thermally-driven wind model therefore predicts that faster solar wind emanates from hotter regions of the corona, and vice versa.
In reality, however, several observations indicate that high-speed solar wind emanates from relatively cool parts of the corona.
Fast solar wind is known to originate from  coronal holes \citep{Krieger_1973SoPh,Kohl_2006A&ARv}, which exhibit cooler temperature than the other regions of the corona \citep[e.g.,][]{Withbroe_1977ARA&A,Narukage_2011SoPh}. 
The observed anti-correlation between the freezing-in temperature and the velocity of the solar wind \citep{Geiss_1995SSRev,von_Steiger_2000JGR} also supports the fact that fast solar wind originates in cool portions of the corona.
These observations indicate that magnetic field plays a substantial role in the solar wind acceleration.

It is believed that the convection beneath the photosphere is the source of the energy for the hot corona and the solar wind \citep{Klimchuk_2006_SolPhys, McIntosh_2011}.
Convective fluctuations excite various modes of waves that propagate upward \citep{Lighthill_1952RSPSA,Stein_1967SoPh,Stepien1988ApJ,Bogdan_1991ApJ}. 
Magnetic reconnection between open and closed field lines is another possible source of transverse waves \citep{Nishuzuka_2008ApJ}, in addition to the direct ejection of heated plasma \citep{Fisk_2003JGRA}.
Among various types of waves, Alfv\'{e}n(ic) waves have been highlighted as a reliable agent to effectively transfer the kinetic energy of the convection to the corona and the solar wind via the Poynting flux \citep[e.g.,][]{Belcher_1971ApJ,Shoda_2019ApJ,Sakaue_2020ApJ,Matsumoto_2021_MNRAS}. This is first because they are not so much affected by the shock dissipation owing to the incompressible nature, unlike compressible waves, which easily steepen to form shocks as a result of the amplification of the velocity amplitude in the stratified atmosphere, and second because they do not refract, unlike fast-mode magnetohydronamical (MHD hereafter) waves \citep[e.g.,][]{Matsumoto_2014MNRAS}, but do propagate along magnetic field lines \citep{Alazraki_1971A&A,Bogdan_2003ApJ}. 

In recent years transverse waves have been detected in the chromosphere \citep{Okamoto_2007Sci,De_Pontieu_2007_Science,McIntosh_2011,Okamoto_2011ApJ,Srivastava_2017NatSR}, whereas it is still under debate whether the sufficient energy required for the formation of the corona and the solar wind propagates into the corona \citep{Thurgood_2014ApJ}.

Once Alfv\'enic waves enter the corona, the key is how the Poynting flux is transferred to the thermal and kinetic energies of the coronal plasma via the dissipation of the waves. Various damping processes of Alfv\'enic waves have been explored, including turbulent cascade \citep{Velli_1989_PhysRevLett,Matthaeus_1999_ApJ,Cranmer_2007ApJS,Verdini_2010_ApJ,Howes_2013PhPl,Perez_2013ApJ,Shiota_2017ApJ,Adhikari_2020ApJS,Zank_2021PhPl}, 
nonlinear mode conversion to compressible waves
\citep{Kudoh_1999ApJ,Suzuki_2005_ApJ,Suzuki_2006_JGRA,Vasheghani_Farahani_2021ApJ}, 
resonant absorption \citep{Ionson_1978_ApJ,VanDoorsselaere2004,Antolin2015_ApJ} and phase mixing \citep{Heyvaerts_1983_AA,DeMoortel2000A&A,Magyar_2017NatSR}.

In contrast to Alfv\'enic waves, acoustic waves have not been considered to be a major player in the coronal heating because the acoustic waves that are excited by $p$-mode oscillations at the photosphere  \citep[e.g.,][]{Lighthill_1952RSPSA,Felipe_2010} rapidly steepen to form shocks before reaching the corona \citep{Stein_1972ApJ,Priest_2014masu.book,Cranmer_2007ApJS} .  
However, \cite{Morton_2019_Nature_Astronomy} pointed out the contribution of $p$-mode oscillations to the generation of Alfv\'enic waves via the mode conversion from longitudinal waves to transverse waves \citep{Cally_2011ApJ}. 
The aim of the present paper is to investigates roles of the acoustic waves that are excited by vertical oscillations at the photosphere in the Alfv\'en wave-drive wind. 
For this purpose, we perform MHD simulations that handle the propagation, dissipation, and mode-conversion of both transverse and longitudinal waves from the photosphere to \shimizu{several tens of solar radii} with self-consistent heating and cooling. 

In Section \ref{sec:method} we describe the setup of our simulations.
We present main results in Section \ref{sec:Results}. 
We discuss related topics in Section \ref{sec:discussion} and summarize the paper in Section \ref{sec:summary}.

\section{Methods} \label{sec:method}
\begin{figure*}[t]
	\label{model}
	\begin{center}
	 \includegraphics[width=18cm]{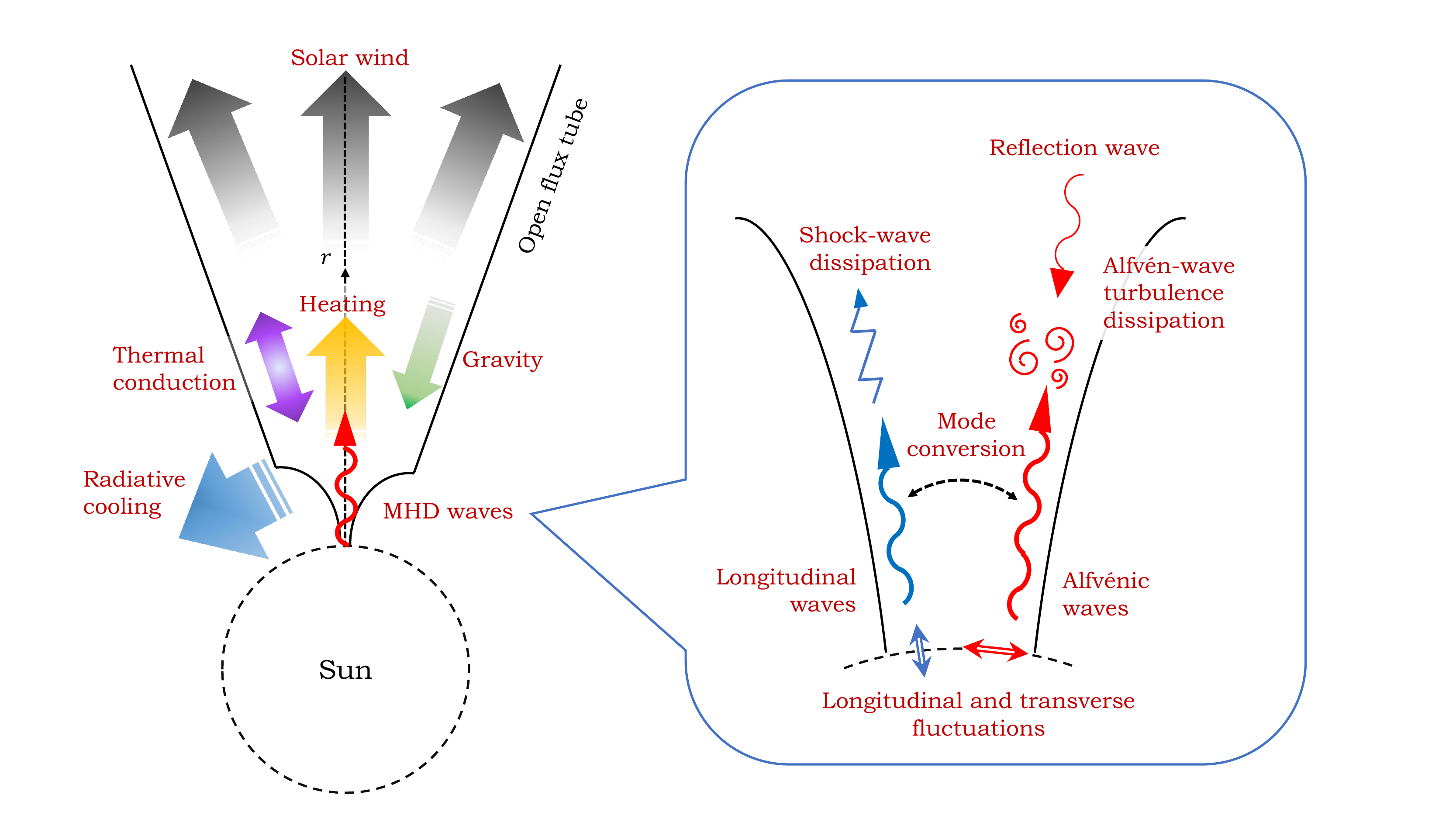}
	 \caption{
	 Schematic pictures of the model.
	 Left picture is the overview of the model.
	 The black solid lines represent the shape of the open flux tube, which expands super-radially. 
	 Right picture in the blue frame is the schematic picture of the wave dissipation in the model.
	 The black dashed curve represents solar surface.
	 Red characters refer to the physical processes considered in this model.
	 \label{fig:model}
	 }
	\end{center}
 \end{figure*}
 
We consider the magnetohydrodynamics of the solar wind in one-dimensional (1D hereafter) open magnetic flux tubes from the photosphere at $r=R_{\odot}$ (solar surface) to  \shimizu{several tens of solar radii}.
Figure \ref{fig:model} shows an overview of our model.

\subsection{Basic Equations}
We consider a one-dimensional (spherical symmetric, $\partial / \partial \theta=\partial / \partial \phi=0$), super-radially expanding flux tube.
The cross section of the flux tube is defined by the filling factor of the open flux tube $f^{\rm op} (r)$, 
which is lower than unity on the photosphere and asymptotically approaches unity as $r$ gets larger.
The conservation of the open magnetic flux $\Phi^{\rm op}$ yields the following relation.
\begin{align}
	\left|B_{r}(r)\right| r^{2} f^{\rm op}(r)=\left|B_{r,\odot}\right|R_{\odot}^2 f^{\rm op}_{\odot}=\Phi^{\rm op},
	\label{eq:divB}
\end{align}
where $X_\odot$ represents the value of $X$ on the photosphere.
We note that $\Phi^{\rm op}$ is constant in each simulation.

We solve the one-dimensional MHD equations along the flux tube characterized by $f^{\rm op} (r)$.
For simplicity, we consider the polar wind, which is not affected by the solar rotation.
In deriving the MHD equations in a super-radially expanding flux tube, the scale factors of the coordinate system are required.
Here, we assume that the magnetic flux tube expands isotropically in $\theta$ and $\phi$ directions.
In terms of scale factors, this assumption yields
\begin{align}
	\label{expf}
	h_{r}=1, \quad h_{\theta}=h_{\phi}=r \sqrt{f^{\rm op}}.
\end{align}
Using these scale factors, the MHD equations in an expanding flux tube is derived (see \cite{Shoda_Takasao_2021arXiv} for derivation).
The basic equations are given as follows.

\begin{align}
	\frac{\partial}{\partial t} \rho + \frac{1}{r^{2} f^{\rm op}} \frac{\partial}{\partial r} \left(\rho v_{r} r^{2} f^{\rm op} \right)=0,
\end{align}
\begin{align}
	&\frac{\partial}{\partial t} \left(\rho v_{r}\right)+\frac{1}{r^2 f^{\rm op}} 
	\frac{\partial}{\partial r}\left[\left(\rho v_r^2 + p_T \right) r^2f^{\rm op}\right] \nonumber \\
	&=-\rho \frac{G M_\odot}{r^2} + 
	\left( \rho \boldsymbol{v}_\perp^2+ 2p \right) \frac{d}{d r} \ln \left( r \sqrt{f^{\rm op}} \right),\label{rovr}
\end{align}
\begin{align}
	&\frac{\partial}{\partial t}\left(\rho \boldsymbol{v}_{\perp}\right)
	+\frac{1}{r^{2} f^{\rm op}} \frac{\partial}{\partial r}\left[\left(\rho v_{r} \boldsymbol{v}_{\perp}-\frac{1}{4 \pi} B_{r} \boldsymbol{B}_{\perp}\right) r^{2} f^{\rm op}\right] \nonumber \\
	&=\left(\frac{B_{r} \boldsymbol{B}_{\perp}}{4 \pi}-\rho v_{r} \boldsymbol{v}_{\perp}\right) \frac{d}{d r} \ln \left(r \sqrt{f^{\rm op}}\right)+\rho \boldsymbol{D}_{v_\perp}^{\text {turb}},
	\label{rovx}
\end{align}
\begin{align}
	\frac{1}{r^{2} f^{\rm op}} \frac{\partial}{\partial r}\left(B_{r} r^{2} f^{\rm op} \right)=0,
\end{align}
\begin{align}
	&\frac{\partial}{\partial t} \bm{B}_{\perp}+\frac{1}{r^{2} f^{\rm op}} \frac{\partial}{\partial r}\left[\left(v_{r} \bm{B}_{\perp}-\bm{v}_{\perp} B_{r}\right) r^{2} f^{\rm op} \right] \nonumber \\
	&=\left(v_{r} \bm{B}_{\perp}-\bm{v}_{\perp} B_{r}\right) \frac{d}{d r} \ln \left(r \sqrt{f^{\rm op}}\right)+\sqrt{4 \pi \rho} \bm{D}_{b_\perp}^{\mathrm{turb}},
	\label{induction_x}
\end{align}
\begin{align}\label{eq:energy}
	&\frac{\partial}{\partial t} e+\frac{1}{r^{2} f^{\rm op}} \frac{\partial}{\partial r} \left[\left(\left(e+p_{T}\right) v_{r}-\frac{B_{r}}{4 \pi}\left(\boldsymbol{v}_{\perp} \cdot \boldsymbol{B}_{\perp}\right) \right) \right. \nonumber r^{2} f^{\rm op}\biggr] \nonumber \\
	&=-\rho v_{r} \frac{G M_{\odot}}{r^{2}} + Q_{\rm C} - Q_{\rm R},
\end{align}
where $\bm{v}$, $\bm{B}$, $\rho$ and $p$ are velocity, magnetic field, density and gas pressure, respectively.
$\bm{v}_\perp$ and $\bm{B}_\perp$ are the perpendicular ($\theta$ and $\phi$) components of $\bm{v}$ and $\bm{B}$, respectively, that is,
\begin{align} 
    \bm{v}_\perp =v_\theta\bm{e}_\theta+v_\phi\bm{e}_{\phi}, \quad \bm{B}_\perp =B_\theta\bm{e}_\theta+B_\phi\bm{e}_{\phi},
\end{align}
where $\bm{e}_\theta$ and $\bm{e}_\phi$ are unit vectors in $\theta$ and $\phi$ direction, respectively.
$M_\odot$ is the solar mass.
$e$ denotes the total energy density per unit volume given by
\begin{align}
	e = e_{\rm int} +\frac{1}{2} \rho \boldsymbol{v}^{2}+\frac{\boldsymbol{B}_{\perp}^{2}}{8 \pi},
	\label{eq:etotal}
\end{align}
where $e_{\rm int}$ is the internal energy density per unit volume.
$p_T$ denotes the total pressure:
\begin{align}   
	p_{T}&=p+\frac{\boldsymbol{B}_\perp^2}{8 \pi}.
	\label{eq:ptotal}
\end{align}    
$\bm{D}_{v_\perp}^{\text {turb }}$ and $\bm{D}_{b_\perp}^{\text {turb }}$ represent the phenomenological turbulent dissipation of Alfv\'en waves (see Section \ref{sec:Alfventurbulence} for detail).

$Q_{\rm C}$ and $Q_{\rm R}$ represent the conductive heating and radiative cooling per unit volume, respectively.
In terms of conductive flux $q_{\rm cnd}$, $Q_{\rm C}$ is given by
\begin{align}
    Q_{\rm C} = - \frac{1}{r^2 f^{\rm op}} \frac{\partial}{\partial r} \left( q_{\rm cnd} r^2 f^{\rm op} \right).
\end{align}
For $q_{\rm cnd}$, 
we employ the Spitzer-Härm type conductive flux \citep{Spitzer_1953_PhRv} that strongly depends on temperature and transports energy preferentially along the magnetic field line.
Besides, to speed up the simulation without loss of reality, we quench the conductivity in the low-density region.
$q_{\rm cnd}$ is then employed as follows.
\begin{equation}
\label{F_C}
	q_{\rm cnd}= - \min \left(1, \frac{\rho}{\rho_{\rm cnd}}\right) \frac{B_{r}}{|\boldsymbol{B}|} \kappa_{0} T^{5 / 2} \frac{d T}{d r}
\end{equation}
where $\kappa_{0}=10^{-6} \operatorname{erg} \mathrm{cm}^{-1} \mathrm{~s}^{-1} \mathrm{~K}^{-7 / 2}$.
We set $\rho_{\rm cnd} = 10^{-20} \mathrm{~g} \mathrm{~cm}^{-3}$, following \citet{Shoda_2020_ApJ}.

The radiative cooling rate is given by a linear combination of optically thick and thin components as follows.
\begin{equation}
	Q_{\rm R} = 
	Q_{\rm R}^{\rm thck} \xi_{\rm rad} + 
	Q_{\rm R}^{\rm thin} \left( 1-\xi_{\rm rad} \right), 
\end{equation}
where $Q_{\rm R}^{\rm thck}$ and $Q_{\rm R}^{\rm thin}$
correspond to the optically thick and thin radiative cooling rates, respectively.
The control parameter $\xi_{\rm rad}$ should satisfy $\xi_{\rm rad} \approx 1$ in the photosphere and $\xi_{\rm rad} \approx 0$ from above the transition region. 
Although the profile of $\xi_{\rm rad}$ is given as a solution of radiative transfer, 
here we simply model it as follows.
\begin{equation}
\label{eq:xi_rad}
	\xi_{\rm rad}=\max\left(0,1-\frac{p_{\rm chr}}{p}\right), 
\end{equation}
where we set $p_{\rm {chr}}=0.1 p_\odot$. 
Thus the radiation is assumed to be optically thick in $p \gtrsim p_{\rm chr}$ and optically thin in $p \lesssim p_{\rm chr}$.

Following \citet{Gudiksen_Nordlund_2005_Apj}, 
we approximate the optically thick cooling by an exponential cooling that forces the local internal energy to approach the reference value $e_{\rm int}^{\rm ref}$:
\begin{align}
	Q_{\rm R}^{\rm thck} = 
	\frac{1}{\tau^{\rm thck}} \left(e_{\rm int} - e_{\rm int}^{\rm ref} \right) ,
	\label{eq:Newtoncl}
\end{align} 
where we set the time scale as follows.
\begin{align}
	\tau_{\rm thck} = 0.1 \left(\frac{\rho}{\bar{\rho}_\odot}\right)^{-1/2} \mathrm{s},
\end{align}
where $\bar{\rho}_\odot=1.87\times10^{-7}\;\mathrm{g}\;\mathrm{cm}^{-3}$, is the mean (time-averaged) mass density in the photosphere. 
The reference internal energy is calculated once the corresponding reference temperature $T^{\rm ref} (r)$ is given.
Here, we set $T^{\rm ref} (r) = T_\odot$.

The optically-thin cooling function is composed of two different contributions.
In the chromospheric temperature range, 
we employ the radiative cooling function given by \cite{Googman_2012ApJ} 
($Q_{\rm GJ}$), 
while in the coronal temperature range,
the loss function $\Lambda (T)$ is given by the CHIANTI atomic database.
\begin{align}
	Q_{\rm R}^{\rm thin} &=Q_{\rm GJ}(\rho, T) \xi_{2} 
	+ n_{\rm H} n_{e} \Lambda(T) \left(1-\xi_{2}\right) ,
\end{align} 
where we set
\begin{align}
	\xi_{2} &=\max \left(0, \min \left(1, \frac{T_{\mathrm{TR}}-T}{\Delta T}\right)\right), \nonumber
\end{align}
where $T_{\text TR}=15000 \ \text K$ and $\Delta T=5000 \ \text K$. 

\subsection{Equation of state}

The hydrogen in  the lower atmosphere (photosphere and chromosphere) of the Sun is partially ionized because the temperature there is not sufficiently high.
In this work, the effect of the partial ionization is considered in the equation of state. 
The internal energy is composed of the random thermal motion of the particles and the latent heat of the hydrogen atoms, which is given by
\begin{align}
    e_{\rm int} = \frac{p}{\gamma-1} + n_{\rm H} \chi I_{\rm H}, \ \ \ \ n_{\rm H} = \rho/m_{\rm H},
\end{align}
where $n_{\rm H}$ is the number density of hydrogen atoms, $\chi$ is the ionization degree and $I_{\rm H} = 13.6 {\rm \ eV}$ is the ionization potential of hydrogen.
For simplicity, the formation of ${\rm H}_2$ molecules is not considered.
The thermal equilibrium is assumed with respect to ionization, in which
the ionization degree is given by the Saha-Boltzmann equation.
\begin{align}
    \frac{\chi^2}{1-\chi} = \frac{2}{n_{\rm H} \lambda_e^3} \exp \left( - \frac{I_{\rm H}}{k_B T} \right),
\end{align}
where $\lambda_e$ is the thermal de Broglie wavelength of an electron:
\begin{align}
    \lambda_e = \sqrt{\frac{h^2}{2 \pi m_e k_B T}}.
\end{align}
Note that pressure and ionization degree are connected by
\begin{align}
    p = \left( 1 + \chi \right) n_{\rm H} k_B T.
\end{align}

\begin{center}
\begin{table*}
\hspace{-15mm}
\scalebox{1.0}{
	\begin{tabular}{ccccccccc}
	\hline \hline
	     Model
		 & $\langle \delta v_{\perp,\odot}^+\rangle$
		 & $\langle \delta v_{\parallel, \odot}\rangle$
		 & $B_{r,\odot}$ 
		 & $f_\odot^{\rm op}$
		 & $f_{\rm chr}^{\rm op}/f_{\rm \odot}^{\rm op}$
		 & $r_\text{out}$
		 & $\dot{M}$
		 & $v_{r,{\rm out}}$ \rule[0mm]{0mm}{4mm} \\
		 &
		 $[{\rm km \ s^{-1}}]$
		 & $[{\rm km \ s^{-1}}]$
		 & $[{\rm G}]$
		 & & & 
		 & $[M_\odot \ {\rm yr}^{-1}]$ 
		 & $[{\rm km \ s^{-1}}]$
		 \rule[-2mm]{0mm}{4mm} \\
	\hline \hline
    B0V06 & 0 & 0.6 & $1.3\times10^{-4}$ & $1.00\times 10^{-3}$ & 100 & $95.6R_\odot$ & accretion & \rule[-2mm]{0mm}{6mm} \\
    \hline
    BsV00 & 0.6 & 0 & 1300 & $1.00\times 10^{-3}$ & 100 & $99.5R_\odot$ & $1.32\times 10^{-14}$ & 688.05 \rule[-2mm]{0mm}{6mm} \\
	BsV04 & 0.6 & 0.4 & 1300 & $1.00\times 10^{-3}$ & 100 & $99.5R_\odot$ & $1.75\times 10^{-14}$ & 687.77 \rule[-2mm]{0mm}{6mm} \\
	BsV06 & 0.6 & 0.6 & 1300 & $1.00\times 10^{-3}$ & 100 & $99.5R_\odot$ & $1.97\times 10^{-14}$ & 697.02 \rule[-2mm]{0mm}{6mm} \\
	BsV09 & 0.6 & 0.9 & 1300 & $1.00\times 10^{-3}$ & 100& $99.5R_\odot$ &$2.63\times 10^{-14}$ & 701.24 \rule[-2mm]{0mm}{6mm} \\
	BsV12 & 0.6 & 1.2 & 1300 & $1.00\times 10^{-3}$ & 100 & $99.5R_\odot$ & $3.10\times 10^{-14}$ & 716.19 \rule[-2mm]{0mm}{6mm} \\
	BsV15 & 0.6 & 1.5 & 1300 & $1.00\times 10^{-3}$ & 100 & $99.5R_\odot$ & $3.54\times 10^{-14}$ & 691.51 \rule[-2mm]{0mm}{6mm} \\
	BsV18  & 0.6 & 1.8 & 1300 & $1.00\times 10^{-3}$ & 100 & $39.1R_\odot$ & $4.18\times 10^{-14}$ & 633.64 \rule[-2mm]{0mm}{6mm} \\
	BsV21 & 0.6 & 2.1 & 1300 & $1.00\times 10^{-3}$ & 100 & $39.1R_\odot$ & $4.57\times 10^{-14}$ & 635.62 \rule[-2mm]{0mm}{6mm} \\
	BsV27 & 0.6 & 2.7 & 1300 & $1.00\times 10^{-3}$ & 100 & $39.1R_\odot$ & $5.09\times 10^{-14}$ & 560.80 \rule[-2mm]{0mm}{6mm} \\
	BsV30 & 0.6 & 3.0 & 1300 & $1.00\times 10^{-3}$ & 100 & $39.1R_\odot$ & $4.97\times 10^{-14}$ & 561.31 \rule[-2mm]{0mm}{6mm} \\
	\hline
	BwV00 & 0.6 & 0 & 325 & $4.00\times 10^{-3}$ & 25 & $95.3R_\odot$ & $1.26\times 10^{-14}$ & 586.55 \rule[-2mm]{0mm}{6mm} \\
	BwV06 & 0.6 & 0.6 & 325 & $4.00\times 10^{-3}$ & 25 & $95.3R_\odot$ & $2.57\times 10^{-14}$ & 581.24 \rule[-2mm]{0mm}{6mm} \\
	BwV18 & 0.6 & 1.8 & 325 & $4.00\times 10^{-3}$ & 25 & $37.8R_\odot$ & $4.76\times 10^{-14}$ & 460.09 \rule[-2mm]{0mm}{6mm} \\
	\hline \hline
	\end{tabular}
	}
	\vspace{0.5em}
	\caption{\label{tab:settings}
	Input parameters (2nd - 7th columns) and output 	results (8th - 9th columns) of different cases.
	}
	\vspace{1em}
\end{table*}
\end{center}

\subsection{Phenomenology of Alfv\'en-wave turbulence}
\label{sec:Alfventurbulence}
In heating the solar wind, energy cascading is required to convert the kinetic and magnetic energies to heat by viscosity and resistivity.
Alfv\'en-wave turbulence, 
a type of MHD turbulence which is triggered by the collision of counter-propagating Alfv\'en waves \citep[e.g.,][]{Goldreich_1995ApJ,Lazarian_2016}, 
is a promising process for the energy cascading in the solar wind.
Because Alfv\'en wave turbulence is a three dimensional process, 
and thus, to deal with the turbulent dissipation in the one-dimensional system,
one needs to model the effect of turbulence.
Here we adopt a phenomenological model of Alfv\'en wave turbulence \citep{Hossain_1995PhFl,Dmitruk_2002,van_Ballegooijen_2016ApJ},
which yields the (averaged) turbulent heating rate in terms of mean-field quantities (Els\"asser variables).
Following \citet{Shoda_2018_ApJ_a_self-consistent_model}, the turbulent dissipation terms in Eq.s (\ref{rovx}) and (\ref{induction_x}) are explicitly given by
\begin{align}
	\label{turblence}
		D_{v_\theta,\phi}^{\text {turb }} &=-\frac{c_{d}}{4 \lambda_{\perp}}\left(\left|z_{\theta,\phi}^{+}\right| z_{\theta,\phi}^{-}+\left|z_{\theta,\phi}^{-}\right| z_{\theta,\phi}^{+}\right) 
\end{align}
and 
\begin{align}
		D_{b_\theta,\phi}^{\text {turb }} &=-\frac{c_{d}}{4 \lambda_{\perp}}\left(\left|z_{\theta,\phi}^{+}\right| z_{\theta,\phi}^{-}-\left|z_{\theta,\phi}^{-}\right| z_{\theta,\phi}^{+}\right),
\end{align}
where $\lambda_\perp$ is a perpendicular correlation length and ${\bm z}_\perp^{\pm}$ are Els\"asser variables \citep{Elsasser_PhRv_1950} defined by
\begin{equation}
	\label{eq:elsasser}
	z_{\theta,\phi}^{\pm}=v_{\theta,\phi} \mp B_{\theta,\phi} / \sqrt{4 \pi \rho} .
\end{equation}
We assume that the correlation length increases with the radius of the flux tube. 
\begin{equation}
	\lambda_{\perp}=\lambda_{\perp, \odot} \frac{r}{R_\odot} \sqrt{\frac{f^{\rm op}}{f_\odot^{\rm op}}}.
	\label{eq:corrlength}
\end{equation}
Because the Alfv\'enic fluctuations are localized in inter-granular lanes on the photosphere \citep{Chitta_2012_ApJ}, we set $\lambda_\perp$ as a typical width of inter-granular lanes. 
\begin{equation}
	\lambda_{\perp, \odot}=150 \ \mathrm{km}
	\label{eq:lambda0}
\end{equation}
The dimensionless coefficient $c_d$ is chosen following \citet{van_Ballegooijen_2017} as
\begin{equation}
	c_{d}=0.1.
\end{equation}

\subsection{Geometory of Flux Tubes}
\label{sec:fluxtube}

In modeling the solar wind in a one-dimensional flux tube, 
we need to prescribe the filling factor of the open flux tube $f^{\rm op}$ as a function of $r$.
Since the open flux tube is localized on the photosphere and expands as the radial distance increases,
$f^{\rm op} (r)$ should be an increasing function of $r$ that asymptotically approaches unity.

Following \citet{Shoda_2020_ApJ}, we employ the two-step expansion of the flux tube,
which is described in terms of $f^{\rm op} (r)$ as 
\begin{align}
	f^{\rm op} (r)=f_{\odot}^{\rm op} f_{1}^{\rm exp}(r) f_{2}^{\rm exp}(r),
\end{align}
where $f_{1}^{\rm exp}(r)$ and $f_{2}^{\rm exp}(r)$ represent the first and second expansions, respectively.

The first expansion occurs in the chromosphere until one flux tube merges with the adjacent flux tube.
Although the direct observation of chromospheric magnetic field is still missing,
because the expansion occurs in response to the exponential decrease in the ambient gas pressure,
it would be straightforward to assume that the filling factor increases exponentially in height.
For this reason, the following formulation is adopted.
\begin{align}
	f_{1}^{\rm exp}(r)=\min \left[f_{\rm cor}^{\rm op} / f_{\odot}^{\rm op}, \exp \left (\frac{r-R_\odot}{H_{\rm mag}} \right)\right],
\end{align}
where $f_{\rm cor}^{\rm op}$ is the open-flux filling factor in the corona and $H_{\rm mag}$ is the scale height of flux-tube expansion.
We relate $H_{\rm mag}$ to the pressure scale height on the photosphere $H_\odot$ by
\begin{align}
    H_{\rm mag} = 2.5 H_\odot = 2.5 \frac{a_\odot^2}{g_\odot},
\end{align}
where $a_{\odot}=6.9 \ {\rm km \; s^{-1}}$ and $g_{\odot}=0.274 \ {\rm km \ s^{-2}}$ are the sound speed and the gravitational acceleration on the photosphere, respectively.

The second expansion occurs in the extended corona with a typical length scale of $\sim R_\odot$.
Following \citet{Kopp_1976_SolPhys},
we adopt the coronal expansion as
\begin{align}
	f_{2}^{\exp }(r)=\frac{\mathcal{F}(r)+f_{\rm cor}^{\rm op}+\mathcal{F}\left(R_{\odot}\right)\left(f_{\rm cor}^{\rm op}-1\right)}{f_{\rm cor}^{\rm op}(\mathcal{F}(r)+1)}, 
\end{align}
where
\begin{align}
    \mathcal{F}(r)=\exp \left(\frac{r-r_{\exp }}{\sigma_{\exp }}\right).
\end{align} 
We adopt the fixed values of $r_{\rm exp} / R_\odot =1.3$, $\sigma_{\rm exp} / R_\odot=0.5$.
The values of $f^{\rm op}_\odot$ and $f^{\rm op}_{\rm cor}$ are summarized in Table \ref{tab:settings}.

\subsection{Simulation Setup}
\label{sec:setup}
The simulation domain extends from the photosphere ($r=R_\odot$) to the outer boundary ($r=r_{\rm out}$) located at nearly $r=100R_{\odot}$ in most cases.
The radial distance of $r_{\rm out}$ for each run is tabulated in Table \ref{tab:settings}.
At $r=r_{\rm out}$, we set the free boundary conditions.
\shimizu{A great advantage to set the inner boundary at the photosphere is that we can self-consistently calculate the density at the coronal base, which 
is one of the critical parameters to determine the mass loss rate, $\dot{M}_w$, of the solar wind \citep[e.g., ][]{Lamers_1999book}. 
The coronal base, where the density is nearly ten orders of magnitude smaller than the density at the photosphere, is frequently set as the inner boundary of simulations for solar and stellar winds \citep{Verdini_2010_ApJ,Lionello_2014_ApJ,Shoda_2019ApJ}.
However, the coronal-base density is determined by the chromospheric evaporation as a result of the energy balance between conductive heating and radiative cooling at the transition region \citep{RTV_1978ApJ,Withbroe_1988ApJ}.
Specifically, when the heating in the corona increases, denser chromospheric material is heated up by the thermal conduction from the corona, resulting in an increase in the density at the coronal base.
Since our numerical simulations solve these heating and cooling processes in a self-consistent manner, we can obtain reliable $\dot{M}_w$ independently from the treatment of the inner boundary at the photosphere.}

\begin{figure}[!t]
	\begin{center}
	 \includegraphics[width=8.5cm]{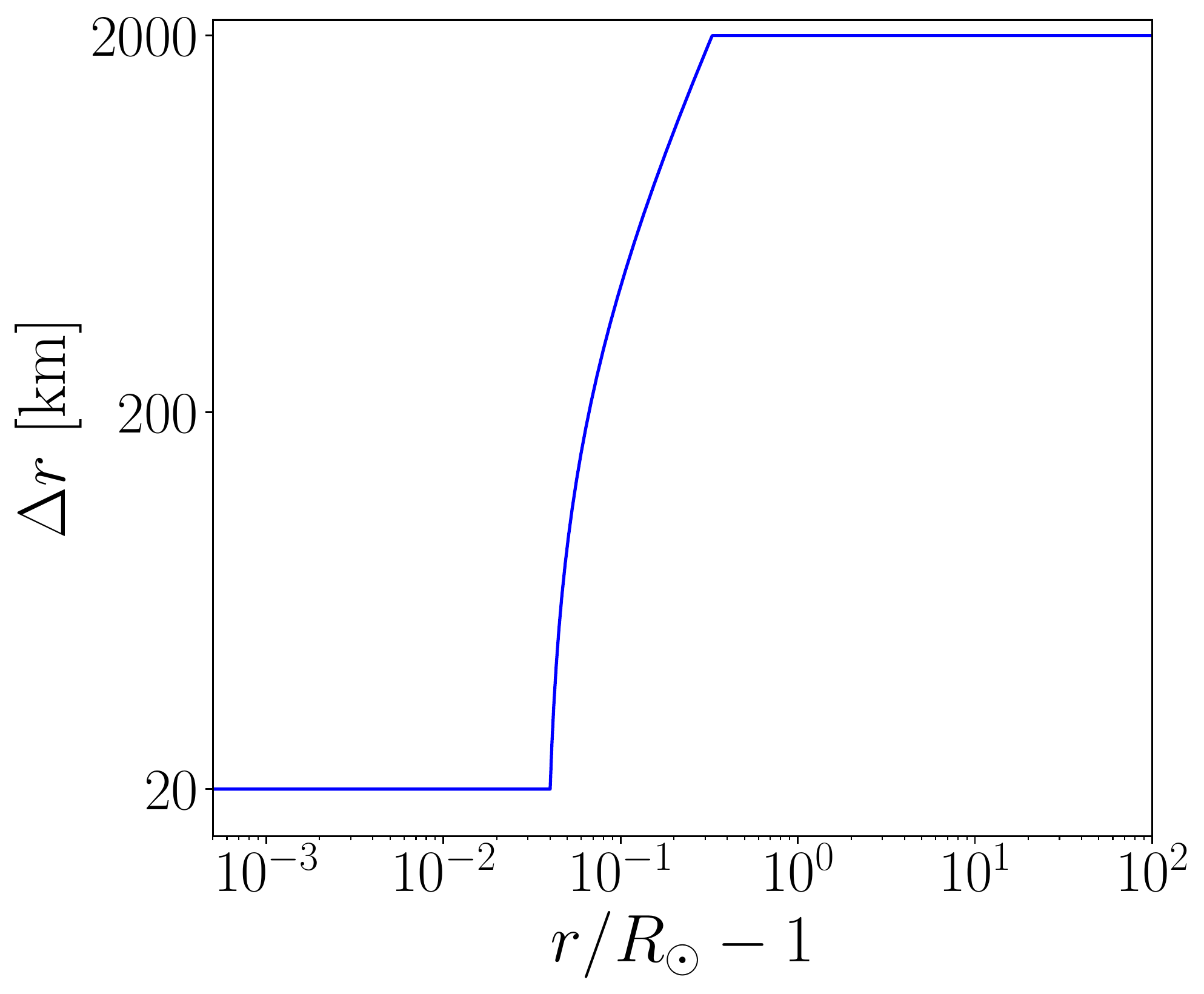}
	 \caption{The radial profile of the grid size, $\Delta r$.  
    \label{grid_size}}
	\end{center}
 \end{figure}

The size of the spatial grid, which varies with $r$,
is set as follows:
\begin{equation}
    \Delta r = \max \left[ \Delta r_{\rm m},{\rm min} \left[ \Delta r_{\rm M},\frac{2 \varepsilon_{\rm ge}}{2 + \varepsilon_{\rm ge}} (r-r_{\rm ge}) +\Delta r_{\rm m} \right] \right]
\end{equation}
where we set $\Delta r_{\rm m}=20 {\rm \ km}$, $\Delta r_{\rm M}=2000 {\rm \ km}$, $\varepsilon_{\rm ge} =0.01$, and $r_{\rm ge} = 1.04 R_\odot$.
Figure \ref{grid_size} shows $\Delta r$ as a function of $r$.

At the inner boundary, we fixed the temperature to the photospheric value, 
\begin{equation}
	T_{\odot} = 5770 \;\mathrm{K} .
\end{equation}
The initial temperature is set to $T=T_{\odot}$ in the entire simulation region. 
We initially set the hydrostatic density distribution with $T=T_{\odot}$ in the inner region that is extended with a power-law profile in the outer region: 
\begin{equation}
\bar{\rho}_{\rm init}(r)=\max \left[ \rho_\odot e^{-\frac{r-R_\odot}{H_{\odot}}},\rho_{\rm w,0}(r/R_\odot-1)^{-2.5}\right],  
\label{eq:initdens}
\end{equation}
where we adopt $\rho_{\rm w,0}=10^{-19}$ g cm$^{-3}$ unless otherwise stated. 
The inner hydrostatic profile switches to the outer power-law one at $r/R_{\odot}-1\approx 0.01$. We note that although the outer density is larger than the hydrostatic value with $T=T_{\odot}$, it is still smaller than the observed density in the solar corona and wind by a factor of five.

The transverse components of velocity and magnetic field correspond to the amplitudes of Alfv\'enic waves. 
The inner boundary condition of them are defined in terms of the Els\"{a}sser variables (Eq. \eqref{eq:elsasser}) in the photosphere.
We set the free boundary condition to the incoming component at the inner boundary so that it is absorbed without being reflected there: 
\begin{equation}
	\left.\frac{\partial}{\partial r} z_{\theta,\phi}^{-}\right|_{\odot}=0
\end{equation}

To inject MHD waves from the photosphere, we impose time dependent boundary conditions for the density, velocity and perpendicular magnetic field. 
The transverse perturbation is injected via the outgoing component of the Els\"asser variables with a broadband spectrum, 
\begin{align}
	z_{\theta,\phi, \odot}^{+} \propto \sum_{N=0}^{100} \sin \left(2 \pi f_{N}^{t} t+\phi_{N}^{t}\right) / \sqrt{f_{N}^{t}},
\end{align}
where $\phi_N^t$ is a random phase and
\begin{align}
	1.00\times10^{-3} \mathrm{Hz} \leq f_{N}^{t} \leq 1.00 \times 10^{-2}\; \mathrm{Hz}.
	\label{eq:freqtransverse}
\end{align}
The longitudinal perturbation, which originates from the $p$-mode oscillation, is excited with the density, 
\begin{align}
		\rho_{\odot}&=\overline{\rho_\odot} \left(1+\frac{v_{r, \odot}}{a_\odot}\right),
\end{align}
where $\overline{\rho_\odot} = 1.88 {\rm \ g \ cm^{-3}}$,
and the radial velocity, 
\begin{align}
		v_{r,\odot}&= \delta v_{\parallel, \odot}(t),
\end{align}
with  
\begin{align}
	\delta v_{\parallel, \odot} \propto \sum_{N=0}^{100} \sin \left(2 \pi f_{N}^{l} t+\phi_{N}^{l}\right) / \sqrt{f_{N}^{l}}, 
\end{align}
where $\phi_N^l$ is a random phase and 
\begin{align}
	3.33 \times 10^{-3} \mathrm{Hz} \leq f_{N}^{l} \leq 1.00 \times 10^{-2}\; \mathrm{Hz}.
\end{align}
This corresponds to the period range between 100 seconds and 5 minutes, which is narrower than that of the transverse component. 

We tabulate the transverse and longitudinal components of the root-mean-squared velocity amplitudes, $\langle \delta v_{\perp,\odot} \rangle$ and $\langle \delta v_{\parallel,\odot} \rangle$, of the input fluctuations at the photosphere in Table \ref{tab:settings}. 
\shimizu{
We take $\langle \delta v_{\perp,\odot}\rangle = \langle \delta v_{\parallel,\odot}\rangle = 0.6$ km s$^{-1}$ as standard values for the velocity perturbation on the photosphere. 
We here note that, because the only outward flux is selectively injected from the photosphere in both transverse and longitudinal components, the corresponding ``random'' velocity amplitude is about $\sqrt{2}$ times larger than these values, which are comparable to observed transverse \citep{Matsumoto_2010ApJ} and longitudinal \citep{Oba2017ApJ} amplitudes of $\sim 1$ km s$^{-1}$.
}

Each model is labeled BxVyy, where ``x'' indicates the type of the magnetic flux tube and ``yy'' denotes the amplitude of $\langle \delta v_{\parallel,\odot} \rangle$. We classify the cases into three groups by the effect of the magnetic field. The first group, which includes only one case, is labeled x $=0$. In this case, we switch off Alfv\'enic waves by setting $\langle \delta v_{\perp,\odot} \rangle = 0$; we test whether the formation of the corona and wind is possible or not only by longitudinal waves. The result of this case is presented in Appendix \ref{sec:sound wave wind} \citep[see also][]{Suzuki_2002ApJ}. 

The second and third groups are labeled x $=$ s and w, which stand for ``standard (or strong)'' and ``weak'' magnetic fields, respectively. 
The aim of these groups is to investigate how the longitudinal-wave excitation on the photosphere affects the properties of the solar wind.
For this purpose, we compare cases with different amplitudes of $\langle \delta v_{\parallel,\odot}\rangle$ for a fixed transverse amplitude of $\langle \delta v_{\perp,\odot}\rangle=0.6$ km s$^{-1}$. 
In the second group, we adopt the equipartition magnetic field, $B_{\odot}=1300$ G, at the photosphere from $8\pi p_{\odot}/B_{\odot}^2=1$ (Section \ref{sec:Results}) to model observed kilo-Gauss patches \citep{Tsuneta_2008ApJ,Ito_2010ApJ}. 
In the third group, in order to examine the effect of the geometry of magnetic flux tubes on the propagation and dissipation of waves in the chromosphere, we reduce $B_{\odot}$ to 1/4 of that of the second group with keeping the field strength above the corona (Section \ref{sec:dependenceonB}).

\shimizu{In order to extensively investigate the effect of longitudinal waves on the solar wind, the second group in particular is investigating a wide range of $0<\langle\delta v_{\parallel,}\odot\rangle<3.0$ km s$^{-1}$. 
$\langle\delta v_{\parallel,}\odot\rangle\gtrsim$2 km s$^{-1}$, which is larger than the observed average value explained above, targets transient large-amplitude disturbances \citep[e.g.,][]{Oba_2017ApJ}. 
}

We perform the simulations for a sufficiently long time in order to study the average behavior of the atmosphere and wind after they reach a quasi-steady state.
To satisfy this requirement, the simulation time is set to 4500 minutes for the cases with $\langle\delta v_{\parallel,\odot}\rangle=0-1.2$ km s$^{-1}$ and  6000 minutes for the cases with $\langle\delta v_{\parallel,\odot}\rangle=1.5-3.0$ km s$^{-1}$. 
Even after the quasi-steady state is achieved, the radial profile fluctuates in time. 
Therefore, when we compare average properties of different cases, we take the average of physical quantities for 1500 minutes before the end of the simulation.

\section{Results}
\label{sec:Results}
In this section, we show results of the cases of the standard magnetic field, BsVyy. 

\subsection{Overview: comparison of radial profiles}

\begin{figure}[ht]
    \begin{center}
    \includegraphics[width=9cm]{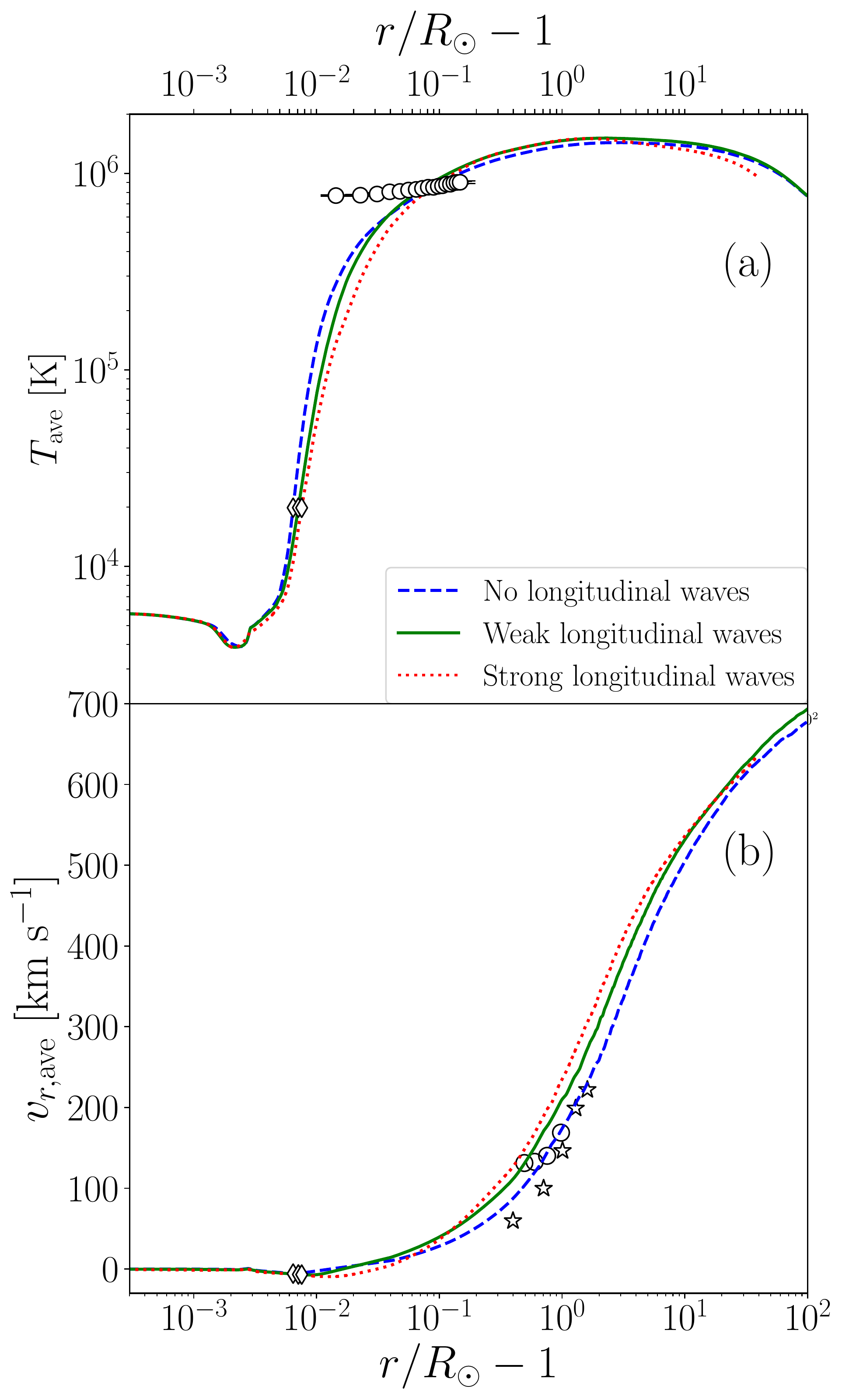}
    \end{center}
    \caption{\label{fig:temp_vr}
    Time-averaged wind profiles of cases with $\langle \delta v_{\parallel,\odot}\rangle=0 \,{\rm km\ s^{-1}}$ (blue dashed; BsV00), $0.6 \,{\rm km\ s^{-1}}$ (green solid; BsV06) and $1.8 \,{\rm km\ s^{-1}}$ (red dotted; BsV18) in comparison with observations. 
	{\bf (a)}: Temperature.
	The circles \citep{Fludra_1999SSRv} show the radial distribution of electron temperature observed by CDS/SOHO.
	{\bf (b)}: Radial velocity.
	            The circles \citep{Teriaca_2003AIPC} and the stars \citep{Zangrilli_2002ApJ} represent  proton outflow speeds in polar regions observed by SOHO.
	The location of the top of the chromosphere at $T = 2\times10^4$ K for each case is shown by diamonds in both panels.
}   
\end{figure}

\begin{figure}[ht]
    \begin{center}
    \includegraphics[width=9cm]{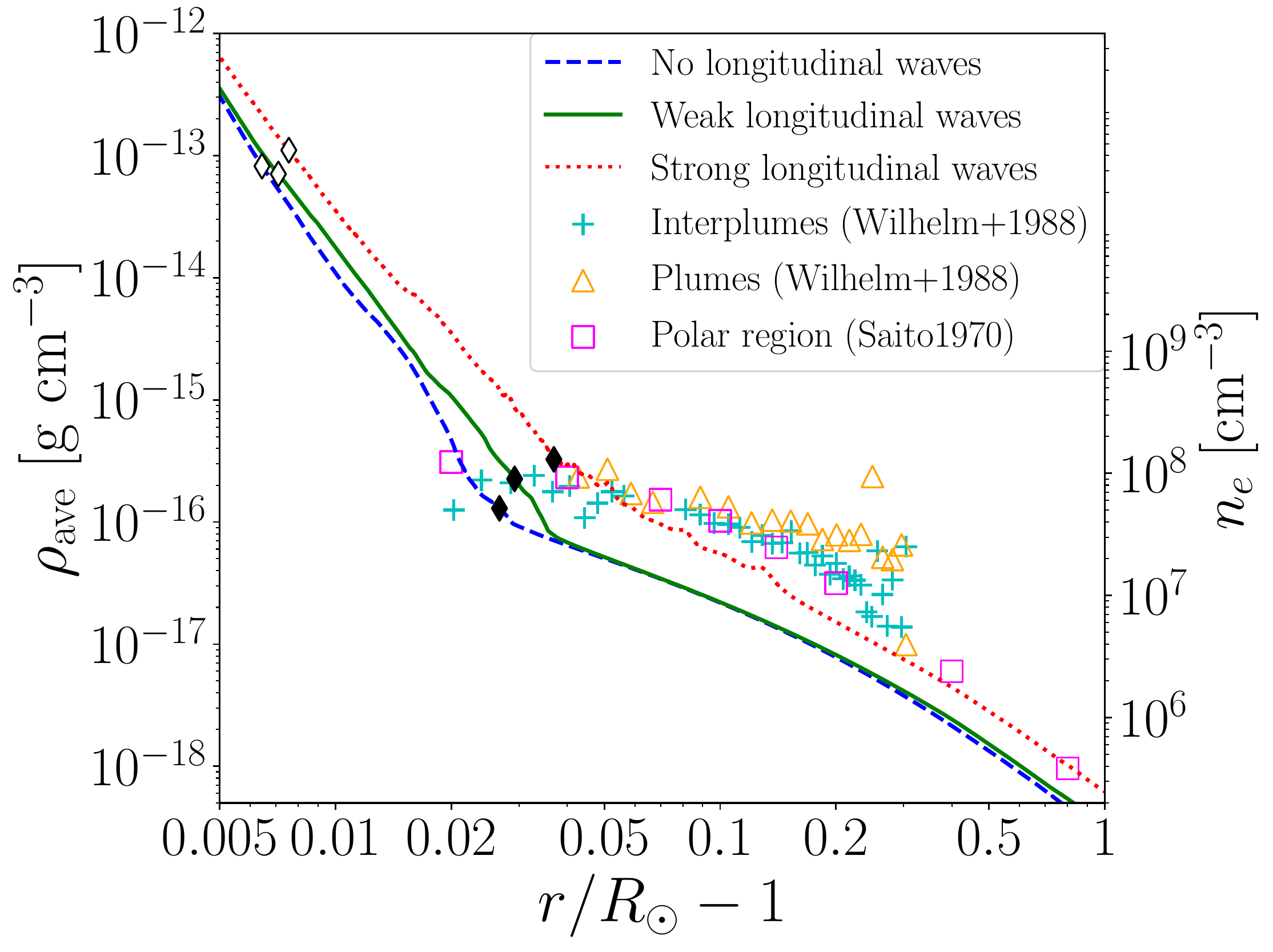}
    \end{center}
    \caption{\label{fig:rho}
    Time-averaged density profiles in the chromosphere and the low corona in comparison to  observations. 
    The line types are the same as those in Figure \ref{fig:temp_vr}.
	The squares represent	electron density (right axis) obtained from observations of multiple total solar eclipses during solar minimum phases \citep{Saito_1970AnTok}.
	The crosses and triangles \citep{Wilhelm_1998ApJ} are electron density observed by SOHO in interplume lanes and plume lanes,  respectively.
	Open and filled diamonds show the location of the top of the chromosphere at $T = 2\times10^4$ K and the location of the coronal base at $T=5\times 10^5$ K, respectively.
}   
\end{figure}

To see the overview, 
we show how the radial profile of the atmosphere and wind depends on the longitudinal-wave amplitude on the photosphere.
Figure \ref{fig:temp_vr} (a) and (b) show the time-averaged radial profiles of the temperature $T$ and radial velocity $v_r$ for three cases:
$\langle \delta v_{\parallel,\odot}\rangle=0.0$ km s$^{-1}$ (BsV00, blue-dashed line),
$\langle \delta v_{\parallel,\odot}\rangle=0.6$ km s$^{-1}$ (BsV06, green-solid line), and
$\langle \delta v_{\parallel,\odot}\rangle=1.8$ km s$^{-1}$ (BsV18, red-dotted line).
Also shown by symbols are the observed values taken from the literature (see the caption for detail).
Several features are found in this comparison.
\begin{enumerate}
    \item The transition region is higher in the large-$\langle \delta v_{\parallel,\odot}\rangle$ cases.
    Given that the upward motion of the transition region (spicules; see Section \ref{sec:TimeVariability}) is likely to be driven by longitudinal waves, 
    the higher transition region is a natural consequence of larger-amplitude longitudinal waves.
    \item No significant differences are seen in the coronal temperature, regardless of the larger energy injection on the photosphere.
    \item In the $v_r$ profiles, while the outflow in the inner region ($r/R_{\odot}-1\lesssim 10$) is slightly faster in large-$\langle \delta v_{\parallel,\odot}\rangle$ cases, the terminal velocity is nearly invariant with $\langle \delta v_{\parallel,\odot}\rangle$.
    This shows that the variety in the solar wind velocity is unlikely to come from the longitudinal-wave injection from the photosphere.
\end{enumerate}

Figure \ref{fig:rho} shows the radial profiles of the mass density $\rho$ (left axis) in the chromosphere and corona ($0.005 \le r/R_\odot-1 \le 1$), 
in comparison to the observed electron densities $n_{\rm e}$ (right axis) in the corona. 
In converting $n_{\rm e}$ to $\rho$,
we assume that the corona is composed of fully ionized hydrogen plasma, that is, $\rho = m_{\rm H} n_{\rm e}$.
The line format is the same as that of Figure \ref{fig:temp_vr}.
In contrast to the temperature and velocity,
the density depends significantly on $\langle \delta v_{\parallel,\odot}\rangle$.
Specifically,
the coronal density is four times larger in $\langle \delta v_{\parallel,\odot}\rangle=1.8$ km s$^{-1}$ than in $\langle \delta v_{\parallel,\odot}\rangle=0.0$ km s$^{-1}$.
Given that the filling factor of the open flux tube $f^{\rm op}$ is fixed and the wind velocity is nearly independent from $\langle \delta v_{\parallel,\odot}\rangle$, 
the larger coronal density means the larger mass-loss rate $\dot{M}_w$, which is given by
\begin{equation}
    \dot{M}_w=4\pi r^2 f^{\rm op}\rho v_r. 
    \label{eq:Mdot}
\end{equation}
Our simulation results show that the wind mass-loss rate is sensitive to the longitudinal-wave injection. 
The underlying physics is discussed in detail in the following sections.

\subsection{Mass-loss rate and Wind Energetics}
\label{sec:MasslossEnergetics}
\begin{figure}[!t]
    \begin{center}
        \includegraphics[width=9cm]{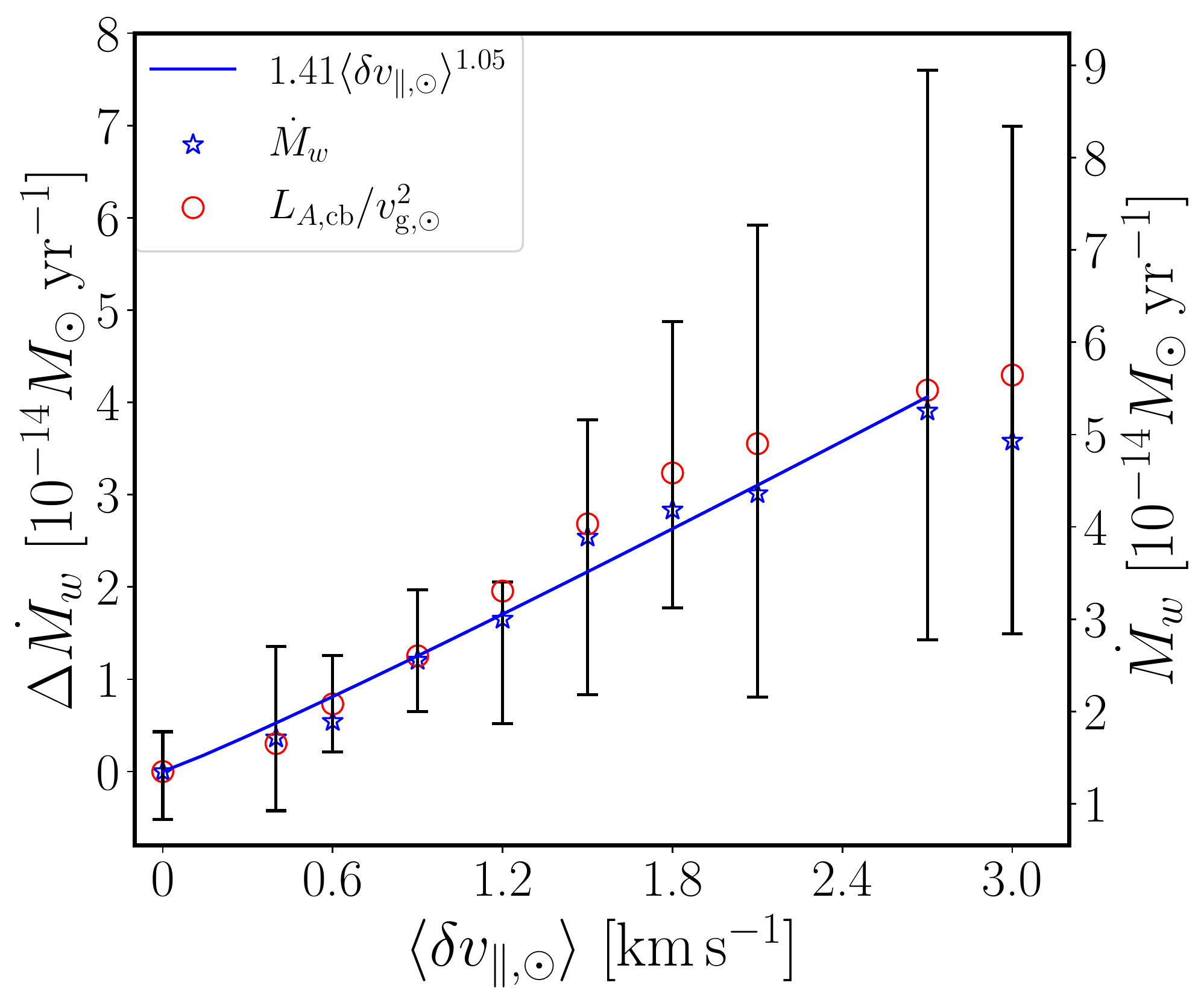} 
    \end{center}
    \caption{\label{fig:vr_mass_loss}
    Mass loss rate versus injected longitudinal wave amplitudes.
     Left and right axes indicate $\Delta \dot{M}_w$ (Equation \eqref{eq:DeltaMdot}) and $\dot{M}_w$ (Equation \eqref{eq:Mdot}), respectively. 
    Blue star symbols with error bars show time-averaged $\dot{M}_w$ of the BsVyy cases with maximum and minimum values during the period of the time average. Red open circles
    represent the theoretical prediction of Equation (\ref{eq:Cranmer_2011}) introduced by \citet{Cranmer_2011_ApJ}. 
    The blue solid line is the power law fit to the time-averaged $\Delta\dot{M}_{w}$ for $\langle \delta v_{\parallel,\odot}\rangle \le 2.7$ km s$^{-1}$(Equation \ref{eq:powerlawfit}).
	 }
\end{figure}

\shimizu{To see more quantitatively how the mass-loss rate depends on the photospheric longitudinal-wave amplitude $\langle \delta v_{\parallel,\odot}\rangle$, 
we present in Figure \ref{fig:vr_mass_loss} the relation between $\langle \delta v_{\parallel,\odot}\rangle$ and the mass-loss rate (blue stars) evaluated at $r=r_{\rm out}$}; shown on the right axis is $\dot{M}_w$ and shown on the left axis is the enhancement of the mass loss rate $\Delta \dot{M}_w$, 
which is defined by
\begin{align}
    \Delta \dot{M}_w = \dot{M}_w - \dot{M}_w^0,
    \label{eq:DeltaMdot}
\end{align}
where $\dot{M}_w^0$ ($=1.32\times 10^{-14}M_{\odot}$ yr$^{-1}$) denotes the mass-loss rate derived from the case with $\langle \delta v_{\parallel,\odot}\rangle = 0.0 {\rm \ km \ s^{-1}}$ (BsV00).
The time variability is also presented by vertical error bars taken from the maximum and minimum values during the time averages. Cases with larger $\langle \delta v_{\parallel,\odot}\rangle$ exhibit higher time variability, which is  discussed later in Section \ref{sec:TimeVariability}. 

The blue solid line in Figure \ref{fig:vr_mass_loss} is the power-law fit to the time-averaged $\Delta \dot{M}_w$ in a range of $\langle \delta v_{\parallel,\odot}\rangle \le 2.7$ km s$^{-1}$: 
\begin{equation}
    \Delta\dot{M}_w=1.41\langle\delta v_{\parallel,\odot}\rangle^{1.05}
    \   10^{-14}M_\odot\; {\rm yr}^{-1}.
    \label{eq:powerlawfit}
\end{equation}
The fitting formula indicates that $\Delta\dot{M}_w$ increases almost linearly with $\langle \delta v_{\parallel,\odot}\rangle$ until saturating above $\langle \delta v_{\parallel,\odot}\rangle \gtrsim 2.7 {\rm \ km \ s^{-1}}$.
The linear dependence indicates that the increase of $\dot{M}_w$ is slower than the increase of the injected energy flux carried by the longitudinal waves $\propto \langle\delta v_{\parallel,\odot} \rangle^2$. 
This implies that not all of the additional input energy of the longitudinal waves, but a portion of it, is used to enhance the mass loss. 
One possible reason is that, as $\langle \delta v_{\parallel,\odot}\rangle$ increases, a larger fraction of the input longitudinal waves
dissipates in the chromosphere due to more efficient shock formation.

\begin{figure}[!t]
    \centering
    \includegraphics[width=8.5cm]{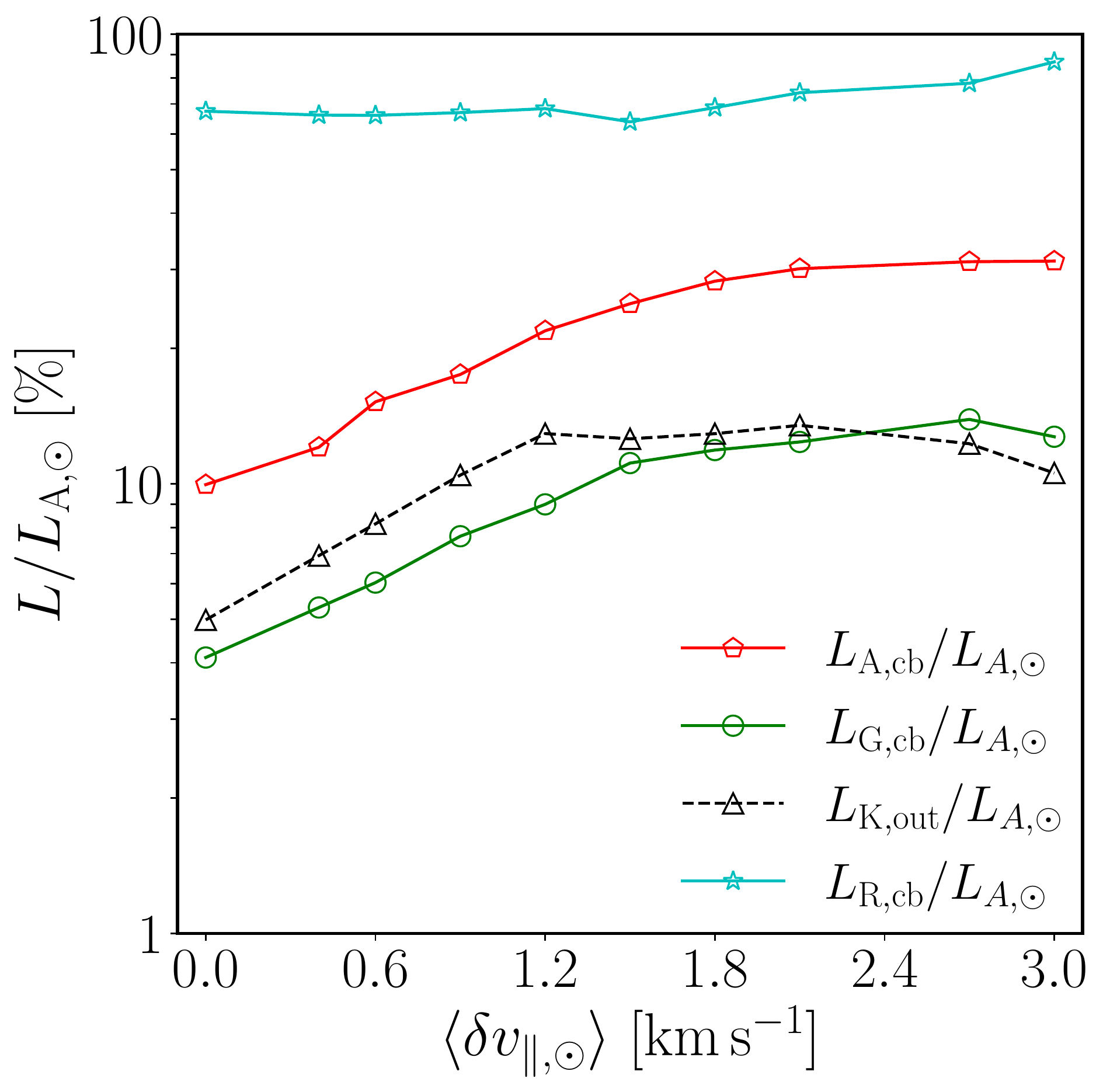}
    \caption{Various components of surface-integrated energy fluxes normalized by the Alfv\'enic Poynting flux at the photosphere with $\langle \delta v_{\parallel_\odot}\rangle$.
    Cian stars, red pentagons, and green open circles with solid lines respectively denote the integrated radiative cooling loss (Equation \eqref{eq:L_R}), Alfv\'enic Poynting flux (Equation \eqref{eq:L_A}), and gravitational potential-energy flux (Equation \eqref{eq:L_G}) measured at the coronal base.
    Black triangles with dashed line represent the kinetic energy flux (Equation \eqref{eq:L_K}) at \shimizu{$r=r_{\rm out}$}.
    \label{fig:energy_flux}
    }
\end{figure}

Although the mass-loss rate depends on the amplitude of the longitudinal wave in the photosphere, it does not mean that the longitudinal wave is the main driver of the solar wind.
As shown in Appendix \ref{sec:sound wave wind},
without transverse wave injection (B0V06),
the atmosphere is heated only up to a few times $10^5$ K and  steady outflows do not occur by the acoustic waves from the photosphere.
Thus, the interaction between longitudinal and transverse waves is possibly the key to understand the cause of the enhancement of the mass loss.
To figure out what caused the increase of $\dot{M}_w$, 
we investigate the global energetics of the wind, 
which is a key to understand the scaling law of mass-loss rate \citep{Cranmer_2011_ApJ,Shoda_2020_ApJ}.
In particular, we consider the radiative energy loss to discuss the energy conservation law from the photosphere to the solar wind \citep{Suzuki_2013_PASJ}.

In the quasi-steady state, the time averaged energy conservation Equation \eqref{eq:energy} is given by
\begin{equation}
	\frac{d}{d r}\left(L_{\rm K}+L_{\rm E}+L_{\rm A}-L_{\rm C}-L_{\rm G}\right) \approx -4 \pi r^{2} f^{\rm op} Q_{\rm R},
	\label{eq:energy_conservation_time_average}
\end{equation}
where 
\begin{align}
    \label{eq:L_K}
	L_{\rm K} &=\frac{1}{2} \rho v_{r}^{3} 4 \pi r^{2} f^{\rm op}, \\
	\label{eq:L_E}
	L_{\rm E} &=\frac{\gamma}{\gamma-1} p v_{r} 4 \pi r^{2} f^{\rm op}, \\
	\label{eq:L_A}
	L_{\rm A} &=\left[\left(\frac{1}{2} \rho \boldsymbol{v}_{\perp}^{2}+\frac{\boldsymbol{B}_{\perp}^{2}}{4 \pi}\right) v_{r}-\frac{B_{r}}{4 \pi}\left(\boldsymbol{v}_{\perp} \cdot \boldsymbol{B}_{\perp}\right)\right] 4 \pi r^{2} f^{\rm op}, \\
	\label{eq:L_C}
	L_{\rm C} &=-q_{\rm cnd} 4 \pi r^{2} f^{\rm op}, \\
	\label{eq:L_G}
	L_{\rm G} &=\rho v_{r} \frac{G M_{\odot}}{r} 4 \pi r^{2} f^{\rm op}=\dot{M}_w \frac{G M_{\odot}}{r} 
\end{align}
are the surface-integrated kinetic energy flux, enthalpy flux, Poynting flux, conductive flux, and gravitational potential-energy flux, respectively.
We note that $\dot{M}_w$ 
in Equation \eqref{eq:L_G} can be assumed to be constant in the quasi-steady state.
We define the radiation luminosity $L_R$ as follows:
\begin{equation}\label{eq:L_R}
    L_{\rm R}(r)= \int_{r_{\rm lch}}^{r}Q_{\rm R} 4\pi r^2f^{\rm op}dr,
\end{equation}
where $r_{\rm lch}$ is the radial distance in the lower chromosphere.
We set $r_{\rm lch}-R_\odot = 0.7 {\rm \ Mm}$ ($r_{\rm lch}/R_{\odot} = 1.001$).
Below $r<r_{\rm lch}$ we assume $L_{\rm R}=0$ because the exponential (Newtonian) cooling, which dominates the radiation in $r \lesssim r_{\rm lch}$, should yield negligible net radiative loss.

Equation \eqref{eq:energy_conservation_time_average} is then rewritten in terms of $L_{\rm R}$ as follows.
\begin{align}
    L_{\rm K}+L_{\rm E}+L_{\rm A}-L_{\rm C}-L_{\rm G} + L_{\rm R} \equiv L_{\rm tot} \approx {\rm const},
    \label{eq:energy_conservation_Ltot}
\end{align}
where $L_{\rm tot}$ is the total surface-integrated energy flux, which is expected to be constant in $r$ in the quasi-steady state.
By relating the values of $L_{\rm tot}$ at different radial distances, several analytical relations are derived.
\begin{enumerate}
    \item 
    Photosphere: Because the kinetic, thermal, and conductive energy fluxes are negligibly small on the nearly static and low-temperature photosphere, 
    the dominant terms in $L_{\rm tot}$ are the Poynting flux and the energy flux of gravitational potential, that is,
    \begin{align}
        L_{\rm tot} \approx L_{{\rm A}, \odot} - L_{{\rm G}, \odot} =  L_{{\rm A}, \odot} - \frac{1}{2}\dot{M}_w v_{g,\odot}^2 ,
        \label{eq:Ltot_photosphere} 
    \end{align}
    where $v_{g,\odot} = \sqrt{2GM_\odot/R_\odot} = 617$ km s$^{-1}$ is the escape velocity.
    We note that $L_{\rm R,\odot}=0$ is assumed as described above. 
    \item
    Coronal base: Because the mean outflow velocity is small at the coronal base, $L_{\rm K}$ and $L_{\rm E}$ are negligible, and thus, $L_{\rm tot}$ is approximated by
    \begin{align}
        L_{\rm tot} \approx L_{{\rm A}, {\rm cb}} - L_{{\rm C}, {\rm cb}} - L_{{\rm G}, {\rm cb}} + L_{{\rm R}, {\rm cb}},
        \label{eq:Ltot_coronal_base_ori}
    \end{align}
    where the subscript ``cb'' denotes the value at the coronal base, which we set $r_{\rm cb}/R_\odot = 1.03$.
    We have confirmed that the conductive luminosity is small at the coronal base because the temperature gradient is already shallow.  
    Therefore, we can safely simplify Equation \eqref{eq:Ltot_coronal_base_ori} to
    \begin{align}
        L_{\rm tot} \approx L_{{\rm A}, {\rm cb}} 
        - L_{\rm G,cb} + L_{{\rm R}, {\rm cb}}. 
        \label{eq:Ltot_coronal_base}
    \end{align}
    \item
    Distant solar wind (outer boundary):
    Because the kinetic energy flux dominates the enthalpy, Poynting, and conductive fluxes in the super-Alfv\'enic region, 
    $L_{\rm tot,out}$ is approximated by
    \begin{align}
        \label{eq:Ltot_solar_wind}
        L_{\rm tot} \approx L_{{\rm K}, {\rm out}} + L_{{\rm R}, {\rm out}} \approx L_{{\rm K}, {\rm out}} + L_{{\rm R}, {\rm cb}},
    \end{align}
    where the subscript ``out'' denotes the value at the outer boundary ($r=r_{\rm out}$).
    Because the radiative loss above the coronal base is generally negligible,
    we use $L_{{\rm R}, {\rm out}} \approx L_{{\rm R}, {\rm cb}}$ (but see discussion below). 
\end{enumerate}

The energy conservation between the photosphere and the coronal base (Eq.s. \eqref{eq:Ltot_photosphere} and \eqref{eq:Ltot_coronal_base}) yields
\begin{align}
    L_{{\rm A}, \odot} \approx L_{{\rm A}, {\rm cb}} + L_{{\rm R}, {\rm cb}},
    \label{eq:LA_coronal_base}
\end{align}
where we approximate $L_{{\rm G},\odot}\approx L_{\rm G,cb}$.
Cian stars and red pentagons in Figure \ref{fig:energy_flux} respectively denote $L_{\rm R,cb}$ and $L_{\rm A,cb}$ normalized by $L_{\rm A,\odot}$. 
Equation \eqref{eq:LA_coronal_base} is satisfied if the sum of these two components is 100 \% in Figure \ref{fig:energy_flux}. 
As one can see, however, this conservation is not perfectly fulfilled, 
possibly because of the treatment of the radiative cooling in the low chromosphere. 
As described previously, $L_{\rm R}$ excludes the contribution of the radiation cooling below $r<r_{\rm lch}$.
By this assumption, we should underestimate $L_{\rm R}$, leading to $L_{{\rm A}, \odot} > L_{{\rm A}, {\rm cb}} + L_{{\rm R}, {\rm cb}}$.

Although we have to bear in mind that $L_{\rm R,cb}$ could be underestimated,
an increasing trend of $L_{\rm R,cb}$ for $\langle \delta v_{\parallel,\odot} \rangle$ is physically plausible; the density in the chromosphere and the low corona is higher for larger $\langle \delta v_{\parallel,\odot} \rangle$ (Figure \ref{fig:rho}), which yields larger radiative cooling. 
As a result, $L_{\rm A,cb}/L_{\rm A,\odot}$ does not monotonically increase with $\langle\delta v_{\parallel,\odot}\rangle$ but eventually saturates for $\langle\delta v_{\parallel,\odot}\rangle \gtrsim 2$ km s$^{-1}$ (Figure \ref{fig:energy_flux}) because a large portion of the input Alfv\'enic Poynting flux is already lost via radiation below the coronal base.

Next, the energy conservation between the coronal base and the outer boundary (Eq.s \eqref{eq:Ltot_coronal_base} and \eqref{eq:Ltot_solar_wind}) yields
\begin{align}\label{eq:L_Kout_approx}
   L_{{\rm K}, {\rm out}} \approx L_{{\rm A}, {\rm cb}}  - L_{\rm G,cb}. 
\end{align}
The green open circles and black triangles in Figure \ref{fig:energy_flux} respectively represent $L_{\rm G,cb}$ and $L_{\rm K,out}$ normalized by $L_{{\rm A},\odot}$. 
$L_{\rm A,cb}$, $L_{\rm G,cb}$ and $L_{\rm K,out}$ in Figure \ref{fig:energy_flux} exhibit a similar trend on $\langle\delta v_{\parallel,\odot}\rangle$; they increase in $\langle\delta v_{\parallel,\odot}\rangle\lesssim 2$ km s$^{-1}$ and saturate for $\langle\delta v_{\parallel,\odot}\rangle\gtrsim 2$ km s$^{-1}$. 
Figure \ref{fig:energy_flux} also shows that Eq.\eqref{eq:L_Kout_approx} is reasonably satisfied, that is, $L_{\rm K,out} + L_{\rm G,cb} \approx L_{\rm A,cb}$.

The wind velocity can be well approximated by the escape velocity, $v_{r,{\rm out}} \approx v_{g,\odot} = \sqrt{2GM_\odot/R_\odot}$, which is also a reasonable assumption in our numerical results (Table \ref{tab:settings}).
Then, by using $L_{\rm K,out}\approx L_{\rm G,cb} \approx \dot{M}_w v_{g,\odot}^2/2$, we can rewrite Equation \eqref{eq:L_Kout_approx} as follows: 
\begin{align}
    \dot{M}_w \approx \frac{L_{{\rm A}, {\rm cb}}}{v_{g,\odot}^2},
    \label{eq:Cranmer_2011}
\end{align}
as already found in \citet{Cranmer_2011_ApJ}.
The comparison of $\dot{M}_w$ (blue stars) to ${L_{{\rm A}, {\rm cb}}}/{v_{g,\odot}^2}$ (orange circles) in Figure \ref{fig:vr_mass_loss} confirms that Equation \eqref{eq:Cranmer_2011} explains the obtained mass loss rate quite well particularly in the small $\langle \delta v_{\parallel,\odot}\rangle \lesssim 1$ km s$^{-1}$ regime. 
In other words, $\dot{M}_w$ is primarily determined by the Alfv\'{e}nic Poynting flux at the coronal base. 
In contrast, Equation \eqref{eq:Cranmer_2011} slightly overestimates the obtained $\dot{M}_w$ in the larger $\langle \delta v_{\parallel,\odot}\rangle \gtrsim 1$ km s$^{-1}$ cases because the radiative cooling above $r>r_{\rm cb}$ is not negligible in Equation \eqref{eq:Ltot_solar_wind} owing to the larger density in the corona (Figure \ref{fig:rho}). 
However, even in these cases with large $\langle \delta v_{\parallel,\odot}\rangle$, Equation \eqref{eq:Cranmer_2011} still gives a reasonable estimate of $\dot{M}_w$.

In summary, the increase and saturation of $\dot{M}_w$ on $\langle \delta v_{\parallel,\odot}\rangle$ directly reflect the trend of the Alfv\'enic Poynting flux at the coronal base. 
The saturation can be interpreted by the excess of the radiative cooling discussed previously. 
On the other hand, in order to understand the increase of $L_{\rm A,cb}$ in $\langle \delta v_{\parallel,\odot}\rangle\le 2.7$ km s$^{-1}$, we need to further examine detailed properties of waves below the coronal base, which is presented in the following subsection. 

\subsection{Dissipation and Mode Conversion of Waves}\label{sec:Alfven energy}
\begin{figure}[!t]
    \begin{center}
        \includegraphics[width=8cm]{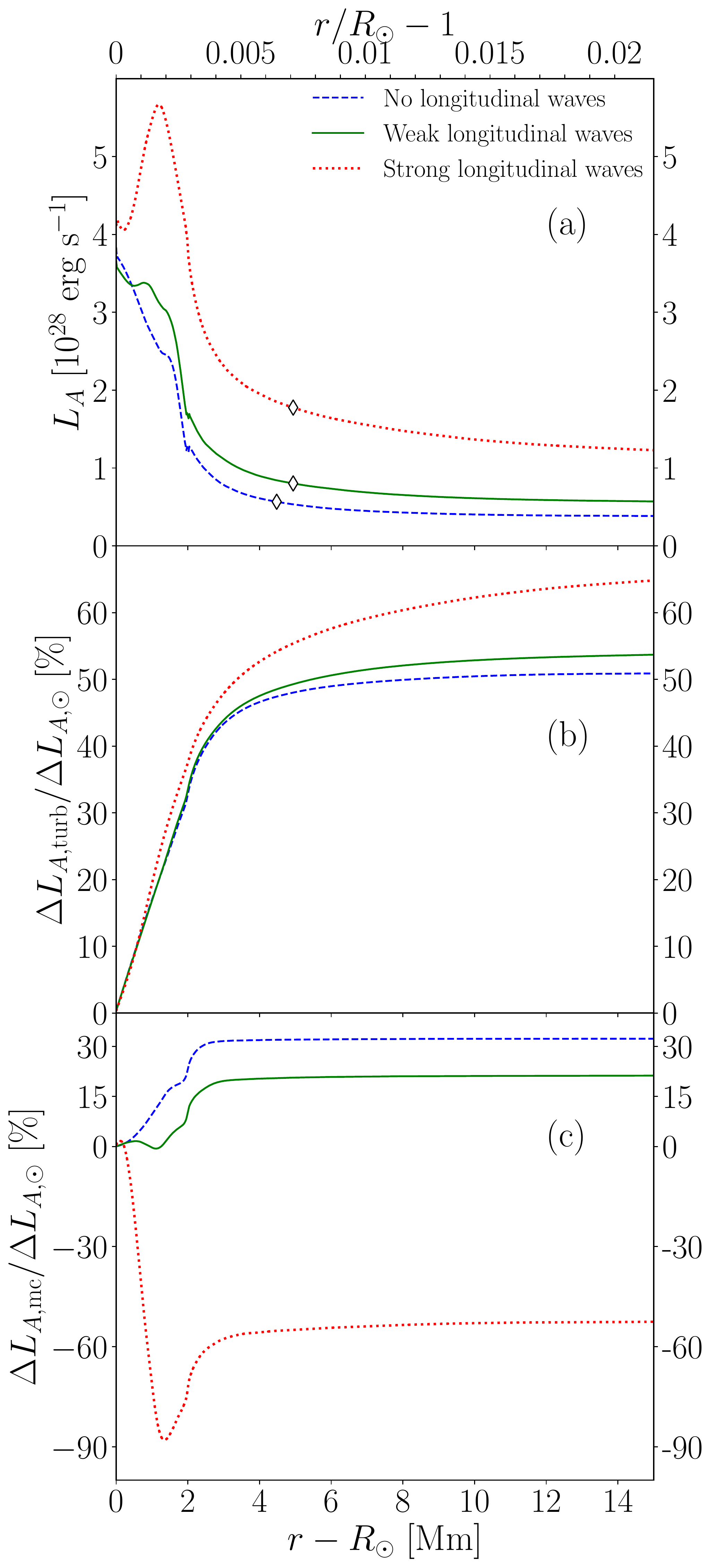}
    \end{center}
    \caption{\label{fig:alfven_energy}
     Comparison of the radial profiles of the surface integrated Alfv\'enic Poynting flux (a) and the fractions of the energy loss by the turbulent dissipation (b) and the mode conversion (c).  
    Blue dashed, green solid and red dotted lines show the results of $\langle \delta v_{\parallel,\odot}\rangle=0 \,{\rm km\ s^{-1}}$ (BsV00), $0.6 \,{\rm km\ s^{-1}}$ (BsV06) and $1.8 \,{\rm km\ s^{-1}}$ (BsV18), respectively.
    In the top panel, the location of the top of the chromosphere at $T = 2\times10^4$ K for each case is shown by diamonds.
    }
\end{figure}

To understand the dependence of $L_{\rm A,cb}$ on $\langle\delta v_{\parallel,\odot}\rangle$ in Figure \ref{fig:energy_flux}, we examine the propagation and dissipation of transverse waves ($\approx$ Alfv\'{e}n waves) from the chromosphere to the low corona. 
\citet{Shoda_2020_ApJ} introduced an equation that describes the evolution of Alfv\'{e}n waves:
 \begin{equation}
	\frac{\partial}{\partial t}\left(\frac{1}{2}\rho v_\perp^2+\frac{B_\perp^2}{8\pi}\right)
	+\frac{1}{4\pi r^2f^{\rm op}}\frac{\partial}{\partial r}L_{\rm A}
	= -\varepsilon_{\parallel\leftrightarrow\perp}-Q_{\rm turb}
\end{equation}
where $\epsilon_{\parallel\leftrightarrow\perp}$ and $Q_{\rm turb}$ indicate the mode conversion from transverse waves to longitudinal waves and the energy loss by turbulent cascade, respectively. 
These are explicitly written as
\begin{align}
    \label{eq:mode_conversion}
	\varepsilon_{\parallel\leftrightarrow\perp}
	&= -v_r\frac{\partial}{\partial r}\left(\frac{B_\perp^2}{8\pi}\right)
	+\left(\rho v_\perp^2-\frac{B_\perp^2}{4\pi}\right)v_r
	\frac{d}{dr}\ln (r\sqrt{f}) \\
	\label{eq:turbulence}
	Q_{\rm turb}
	&= c_{d} \rho \sum_{i=\theta, \phi} \frac{\left|z_{i}^{+}\right| (z_{i}^{-})^2+\left|z_{i}^{-}\right| (z_{i}^{+})^2}{4 \lambda_{\perp}}.
\end{align}
We note that the first term of Equation (\ref{eq:mode_conversion}) denotes the nonlinear excitation of longitudinal perturbation from the magnetic fluctuation associated with transverse waves \citep{Hollweg_1971_JGR,Kudoh_1999ApJ,Suzuki_2005_ApJ,Matsumoto_2010_ApJ}. 
Using $\varepsilon_{r\leftrightarrow\perp}$ and $Q_{\rm turb}$, we define the the energy loss rates via the turbulent dissipation and the mode conversion as
\begin{align}
	\Delta L_{\mathrm{A}, \mathrm{turb}}(r) &=\int_{R_{\odot}}^{r} d r 4 \pi r^{2} f^{\mathrm{op}} Q_{\mathrm{turb}} 
\end{align}
and
\begin{align}
	\Delta L_{\mathrm{A}, \mathrm{mc}}(r) &=\int_{R_{\odot}}^{r} dr 4 \pi r^{2} f^{\mathrm{op}} \varepsilon_{\parallel \leftrightarrow \perp}.
\end{align}

Figure \ref{fig:alfven_energy} shows properties of the damping of Alfv\'enic waves in the chromosphere; Panel (a) presents the radial profile of $L_{\mathrm{A}}$;
Panels (b) and (c) present $\Delta L_{\mathrm{A}, \mathrm{turb}}$ (energy loss by turbulence) and $\Delta L_{\mathrm{A}, \mathrm{mc}}$ (energy loss by mode conversion), respectively.
We note that the net loss of Alfv\'enic waves, $L_{\mathrm{A},{\odot}}-L_\mathrm{A}$, is not always equal to the sum of these energy losses, possibly because of numerical dissipation.

As shown in Figure \ref{fig:alfven_energy}, the mode conversion rate and the turbulent loss rate behave differently.
A general trend is that $\Delta L_{\mathrm{A}, \mathrm{turb}}$ increases and $\Delta L_{\mathrm{A}, \mathrm{mc}}$ decreases with increasing $\langle \delta v_{\parallel,\odot}\rangle$.
As $\langle \delta v_{\parallel,\odot}\rangle$ increases, the mode conversion from transverse (Alfv\'{e}nic) waves to longitudinal (acoustic) waves is suppressed; instead, the ``inverse conversion'' (longitudinal-to-transverse wave energy transfer), $\Delta L_{\mathrm{A}, \mathrm{mc}}<0$, takes place in the case with large $\langle \delta v_{\parallel,\odot}\rangle$ (red dotted; BsV18). 
This is because transverse waves are excited at the region with plasma $\beta\approx 1$ in the chromosphere by the mode conversion from the large-amplitude longitudinal waves injected from the photosphere \citep{Cally_2006, Schunker_2006, Cally_2008}. 
We here note that the conversion from the longitudinal mode to the transverse mode occurs even in the simple 1D geometry because 
the direction of magnetic field, $(B_r\hat{\boldsymbol{r}}+\boldsymbol{B}_{\perp})/|B|$, is not parallel with the direction of the wave propagation that is strictly along the $r$ direction, where $\hat{\boldsymbol{r}}$ is the radial unit vector.

As a result of the excitation of transverse wave from longitudinal wave, 
$L_{\rm A}$ increases near the surface (Figure \ref{fig:alfven_energy}(a)). 
The amplitude of the excited transverse waves increases with $\langle\delta v_{\parallel,\odot}\rangle$, which raises the Alfv\'enic Poynting flux at the coronal base (Figure \ref{fig:energy_flux}). 
\shimizu{
As a consequence of increased energy injection to the corona (increased $L_{\rm A,cb}$), the coronal heating rate increases, which leads to larger coronal density  (see discussion in Section \ref{sec:setup}).
In fact, as shown in Figure \ref{fig:rho}, the density at the coronal base (where $T=5\times 10^5$ K) is higher for larger $\langle\delta v_{\parallel,\odot}\rangle$, even though the coronal base is located at a higher altitude.
}

\begin{figure}[!t]
    \begin{center}
        \includegraphics[width=8cm]{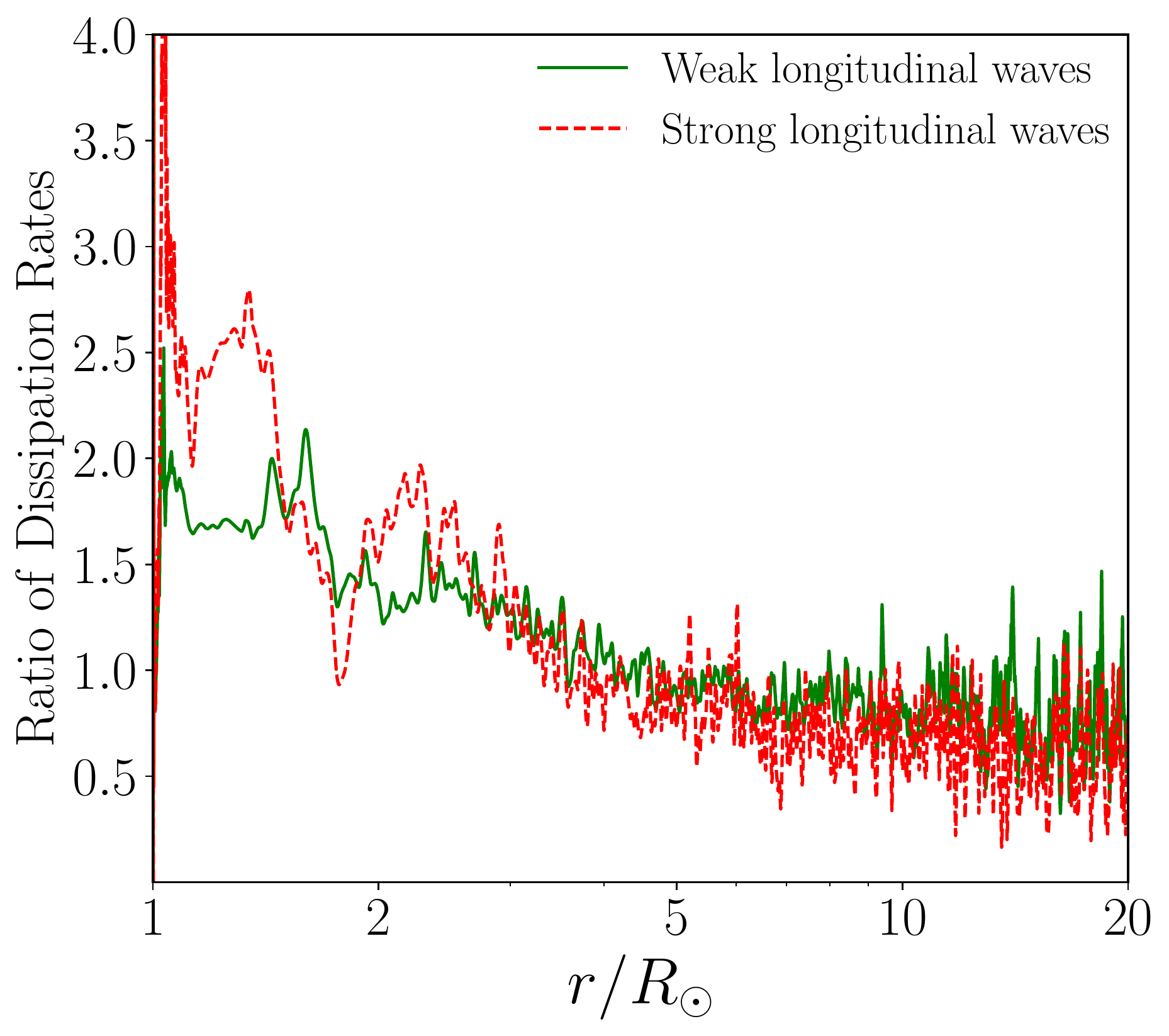}
    \end{center}
    \caption{\label{fig:Alfven_critical}
    \shimizu{Dissipation rate of Alfv\'{e}nic Poynting flux per unit mass (Equation \eqref{eq:dissipationrate}) of $\langle \delta v_{\parallel,\odot}\rangle=0.6 \,{\rm km\ s^{-1}}$ (BsV06; green solid line) and $1.8 \,{\rm km\ s^{-1}}$ (BsV18; red dashed line). The values are normalized by that of $\langle \delta v_{\parallel,\odot}\rangle=0$ (BsV00).}
    }
\end{figure}

\shimizu{
Another interesting point is that the velocity of the wind is insensitive to the value of $\langle \delta v_{\parallel,\odot}\rangle$ (bottom panel of Figure \ref{fig:temp_vr}); the increase of $\dot{M}_w (\propto \rho v_r)$ is solely by the increase in the density. 
According to the standard model of the solar/stellar winds \citep{Hansteen_1995_JGR,Lamers_1999book}, the additional heating and momentum inputs in the subsonic region ($v_r<a$) of a wind raise the mass-loss rate with negligible effects on the terminal velocity, while those in the supersonic region ($v_r>a$) do not affect the mass-loss rate but result in the higher terminal velocity.}
\shimizu{Based on this background, to understand the behavior of wind speed with respect to $\langle \delta v_{\parallel,\odot}\rangle$, we examine the radial distribution of the energy transfer rate from the Alfv\'enic wave to the gas from the corona to the wind.}
\shimizu{Specifically, we calculate the loss rate of the Alfv\'{e}nic Poynting flux per unit mass, defined as
\begin{align}
    \zeta_{\rm A}=-\frac{1}{4\pi\rho r^2 f^{\rm op}}\frac{\partial L_{\rm A}}{\partial r}.
    \label{eq:dissipationrate}
\end{align}
Because of energy conservation, $\zeta_{\rm A}$ corresponds to the energy ($=$ heating $+$ work) transfer rate from the Alfv\'{e}nic Poynting flux to the plasma.
}
\shimizu{Figure \ref{fig:Alfven_critical} presents $\zeta_{\rm A}$ of the cases with $\langle \delta v_{\parallel,\odot}\rangle = 0.6$ km s$^{-1}$(BsV06; green solid) and 1.8 km s$^{-1} $(BsV18; red dashed) normalized by $\zeta_{\rm A}$ of $\langle \delta v_{\parallel,\odot}\rangle = 0$ (BsV00). }
\shimizu{The increase in $\langle \delta v_{\parallel,\odot}\rangle$ promotes the energy input in the subsonic region ($r\lesssim 2-3R_{\odot}$) but does not affect (or even reduces) the energy input in the supersonic region ($r\gtrsim 2-3R_{\odot}$). 
In other words, the vertical oscillation on the photosphere affects only the subsonic region.
For this reason, an addition of $\langle \delta v_{\parallel,\odot}\rangle$ does not affect the wind velocity but only increases the mass-loss rate.
}

\section{Discussion}
\label{sec:discussion}

\subsection{Dependence on Magnetic Field in Chromosphere}
\label{sec:dependenceonB}
\begin{figure}[h]
    \begin{center}
    \includegraphics[width=8cm]{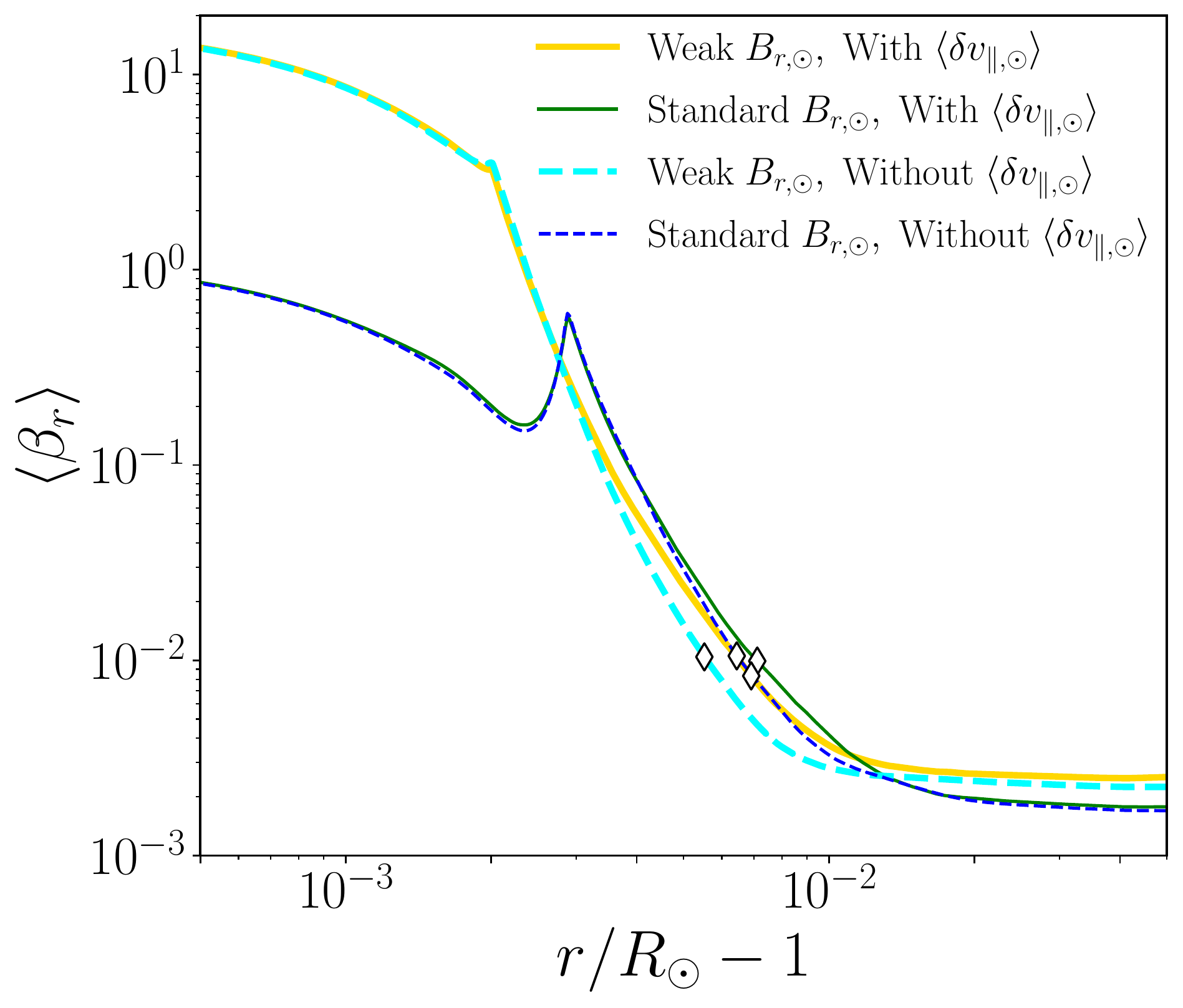}
    \end{center}
    \caption{\label{fig:beta}
    Comparison of the radial profiles of time averaged $\langle \beta_r\rangle$ (Equation \eqref{eq:betar}). 
    Thin and thick lines correspond to the cases of $B_{r,\odot}=1300$ G (BsVyy) and $325$ G (BwVyy), respectively.
    Dashed and solid lines show the cases with $\langle \delta v_{\parallel,\odot}\rangle = 0$ (BxV00) and 0.6 km s$^{-1}$ (BxV06), respectively. 
	Diamonds show the location of the top of the chromosphere at $T = 2\times10^4$ K for each case.
   }
\end{figure}

We have shown that the nonlinear mode conversion between transverse waves and longitudinal waves in the chromosphere is the key to determine the wind properties when both transverse and longitudinal perturbations are input at the photosphere.
The mode conversion rate sensitively depends on plasma $\beta=8\pi p/B^2$ with peaked at $\beta\approx 1$ \citep{Hollweg_1982_ApJ,Spruit_1992_ApJ}.
Therefore, it is expected that the magnetic field strength in the chromosphere plays an essential role in determining the Alfv\'{e}nic 
Poynting flux that enters the corona. 
Here, we perform simulations in a flux tube with weaker magnetic field from the photosphere to the chromosphere but with the same field strength above the corona (``BwVyy'' in Table \ref{tab:settings}).

Figure \ref{fig:beta} compares the radial profiles of the time averaged plasma beta values of four cases, BsV00, BsV06, BwV00, and BwV06, that are evaluated from the only radial component of the magnetic field, 
\begin{align}
    \langle\beta_r\rangle \equiv \frac{8\pi \langle p\rangle}{B_r^2}. 
    \label{eq:betar}
\end{align}
We note that, although $\beta_r \ge \beta$, the difference is small because $|\delta B_{\perp}|<|B_r|$. 

\begin{figure}[t]
    \begin{center}
    \includegraphics[width=8cm]{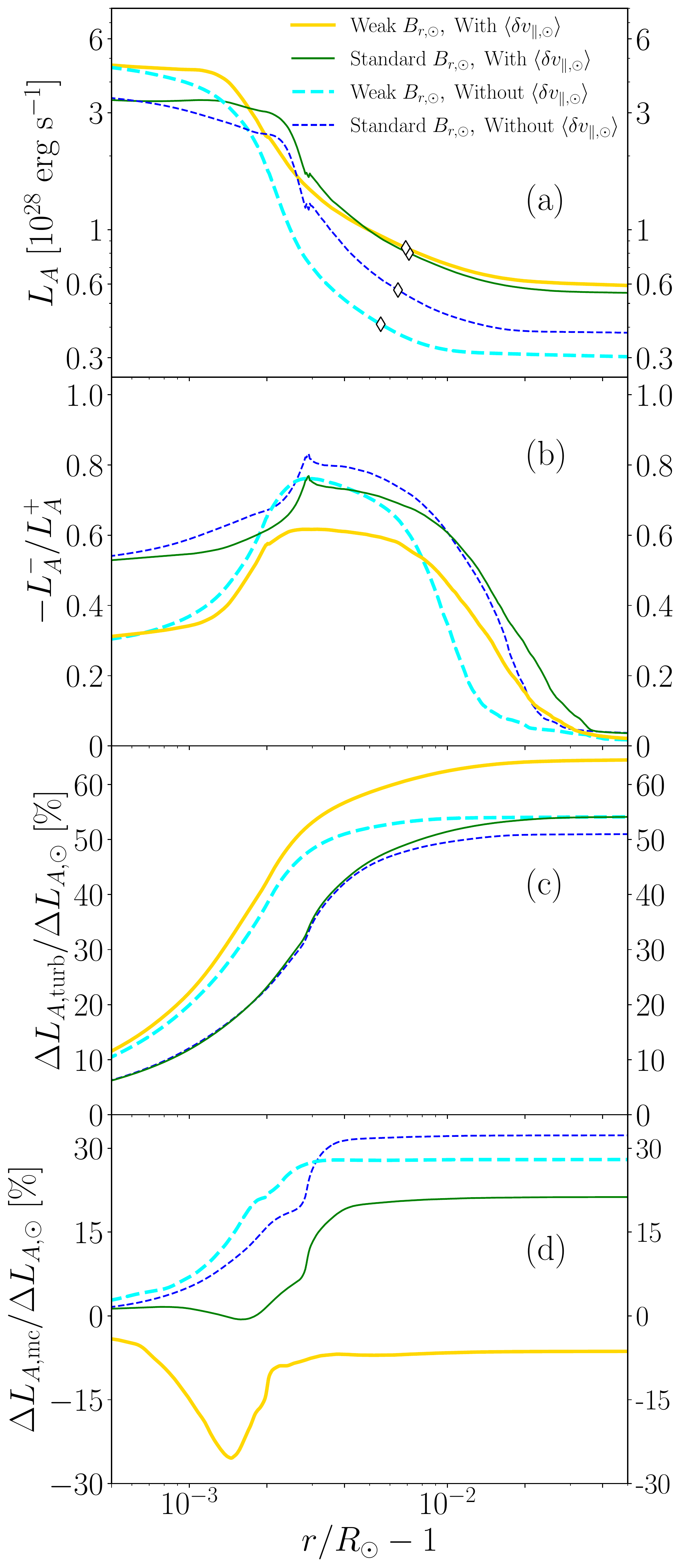}
    \end{center}
    \caption{\label{fig:alfven_energy2}
    Comparison of the radial profiles of the surface-integrated 
    Alfv\'enic Poynting flux (a), the ratio of the inward Poynting flux to the outward Poynting flux
    (b), and the fraction of the energy loss by the turbulent dissipation (c) and the mode conversion (d).
    The line types are the same as those in Figure \ref{fig:beta}. 
    In the top panel, the location of the top of the chromosphere at $T = 2\times10^4$ K for each case is shown by diamonds.
    }
\end{figure}
Figure \ref{fig:alfven_energy2} compares the properties of Alfv\'{e}n waves of these four cases. The panels (a), (c), and (d) are the same as (a), (b) and (c) of Figure \ref{fig:alfven_energy} but the vertical axis of (a) and the horizontal axis are shown in logarithmic scale. The panel (b) presents the ratio of the incoming component $L_{\rm A}^{-}$, to the outgoing component $L_{\rm A}^{+}$, of Alfv\'{e}nic Poynting luminosity,
which are defined by
\begin{align}
    L_{\rm A}^{+} &= \rho (z_\perp^{+})^2 (v_r + v_{\rm A}) \pi r^2 f^{\rm op}, \\
    L_{\rm A}^{-} &= \rho (z_\perp^{-})^2 (v_r - v_{\rm A}) \pi r^2 f^{\rm op},
\end{align}
where $v_{\rm A}=B_r/\sqrt{4\pi\rho}$ is the Alfv\'en velocity along the $r$ direction.
We note that $L_{\rm A} = L_{\rm A}^{+} + L_{\rm A}^{-}$. 

Let us begin with the comparison of the cases without $\langle \delta v_{\parallel,\odot}\rangle$, BwV00 (light blue thick dashed lines) and BsV00 (deep blue thin dashed lines). 
Figure \ref{fig:alfven_energy2} (a) shows that $L_{\rm A}$ of the weak field case (BwV00) is larger than $L_{\rm A}$ of the standard field case (BsV00) near the photosphere. 
However, 
the former declines more rapidly in the chromosphere to be 
smaller than 
the latter above the upper chromosphere of $r/R_{\odot}-1>2\times 10^{-3}$. 
As a result, the mass loss rate $\dot{M}_w$ of BwV00 is slightly smaller than $\dot{M}_w$ of BsV00 (Table \ref{tab:settings}), which is consistent with Equation \eqref{eq:Cranmer_2011}.
The rapid damping of the Alfv\'en waves is mainly because of more efficient turbulent dissipation (Figure \ref{fig:alfven_energy2}(c)). 
Utilizing Equation \eqref{eq:divB}, we can rewrite the correlation length (Equation \eqref{eq:corrlength}) as follows:
\begin{equation}
    \lambda_\perp=\lambda_{\perp,\odot}\frac{r}{R_\odot}\sqrt{\frac{f^{\rm op}}{f^{\rm op}_\odot}},  
\end{equation}
where we are adopting the same $\lambda_{\perp,\odot} = 150 {\rm \ km}$ 
in both cases (Equation \eqref{eq:lambda0}). 
Since the flux-tube expansion of the weak field case is smaller in our setup, $f^{\rm op}/f^{\rm op}_\odot$ is smaller, which enhances the turbulent dissipation.
The rapid turbulent damping in the chromoshere reduces the amplitude of the outgoing Alfv\'{e}n waves in the upper region. 
The reflected Alfv\'{e}n waves downward to the photosphere are also suppressed to give the smaller ratio of $L_{\rm A}^{-}/L_{\rm A}^{+}$ near the photosphere of the weak field case (Figure \ref{fig:alfven_energy2}(b)). 
Therefore, the net outgoing Poynting flux, $L_{\rm A} (=L_{\rm A}^{+} + L_{\rm A}^{-})$, is larger there (Figure \ref{fig:alfven_energy2}(a)). 

The comparison of BwV06 to BwV00 indicates that $\dot{M}_w$ increases more than twice by the additional input of the longitudinal perturbation of $\langle \delta v_{\parallel, \odot}\rangle = 0.6$ km s$^{-1}$ in the weak field condition (Table \ref{tab:settings}). 
The enhancement factor of $\dot{M}_w$ is considerably larger than the value ($\approx 1.5$ times) obtained in the standard field condition. 
This is because in the weak field case of BsV06 (orange thick lines) larger Alfv\'enic Poynting flux reaches the coronal base (Figure \ref{fig:alfven_energy2}(a)) by the generation of transverse waves through the mode conversion (Figure \ref{fig:alfven_energy2}(d)) in spite of the higher turbulent loss (Figure \ref{fig:alfven_energy2}(c)).
Figure \ref{fig:beta} shows that $\langle \beta_r\rangle$ of this case decreases with height and crosses unity in the chromosphere, which induces the efficient longitudinal-to-transverse mode conversion (Figure \ref{fig:alfven_energy2}(d)) as shown by \citet{Cally_2006} and \citet{ Cally_2008}. 
In contrast, $\langle\beta_r\rangle$ stays $< 1$ in the standard field case of BsV06 (Figure \ref{fig:beta}). 
As a result, $\Delta L_{\rm A,mc}$ remains positive (Figure \ref{fig:alfven_energy2} (d)), namely the excitation of transverse waves by the mode conversion is negligible. 
We can conclude that, in addition to the longitudinal fluctuation in the photosphere (Section \ref{sec:Results}), the magnetic field strength in the chromosphere is also an essential factor to determine the global wind properties through nonlinear processes of MHD waves. 

\subsection{Limitation of the 1D Geometry}

We have simulated the propagation, dissipation, and mode conversion of MHD waves in 1D super-radially open flux tubes. 
While we took the phenomenological approach to the Alfv\'en-wave turbulence (Section \ref{sec:Alfventurbulence}) to consider the 3D effect, multi-dimensional effects are also important in other wave processes \citep[e.g.][]{Hasan_2008ApJ,Matsumoto_2012ApJ,Iijima_2017ApJ,Matsumoto_2021_MNRAS}.
The nonlinear mode conversion, which is a key process in the present paper, is probably one of those that have to take into account 3D effects because the conversion rate increases with the attacking angle between the direction of a magnetic field line and a wave-number vector \citep{Schunker_2006}. 
Since the attacking angle tends to be restricted to a small value in the 1D treatment, the amount of the generated transverse waves by the mode conversion may be underestimated in our simulations. 

Although we have only considered shear Alfv\'{e}n waves, torsional Alfv\'{e}n waves are also expected to be excited \citep{Kudoh_1999ApJ}. The nonlinear steepening of the torsional mode is slower than that of the shear mode  \citep{Vasheghani_Farahani_2012A&A}. 
Therefore, if we considered torsional waves in addition to shear waves, the dissipation rate of the transverse waves would be slower, which may affect the global wind properties.

\subsection{Missing physics in the chromosphere}
We described that  the radiative cooling in the chromosphere governs the saturation of the Alfv\'enic Poynting flux that reaches the coronal base in the cases of the large 
$\langle\delta v_{\parallel,\odot}\rangle$ (Figure \ref{fig:energy_flux} and Sections \ref{sec:MasslossEnergetics} \& \ref{sec:Alfven energy}). 
The local thermodynamical equilibrium is not strictly satisfied in the chromosphere, and the radiative cooling is governed by multiple bound-bound transitions \citep{Carlsson_2012A&A}.  
In addition, the radiative loss also affects the propagation of compressional waves such as acoustic waves \citep[e.g.,][]{Bogdan_1996ApJ}. 
Ideally, detailed radiative transfer has to be solved to accurately handle these complicated processes, although we have taken the approximated prescription to consider them phenomenologically (Section \ref{sec:MasslossEnergetics}). 
A more accurate treatment \citep[e.g.,][]{Hansteen2015ApJ,Iijima_2017ApJ} might modify the radiative loss rate, which we plan to tackle in our future works. 

The gas in the chromosphere is partially ionized plasma. 
The relative motion between charged particles and
 neutral particles, which is called ambipolar diffusion, 
promotes damping of transverse waves and heating the gas \citep{Khodachenko2004A&A,Khomenko_2012ApJ}. 
However, in the current condition of the solar chromosphere, the ambipolar diffusion does not give a large impact on the low-frequency Alfv\'en waves with $<10^{-2}$ Hz considered in this paper \citep[Equation \eqref{eq:freqtransverse};][]{Arber2016ApJ}, although it may affect higher-frequency Alfv\'en waves and rapid dynamical phenomena \citep[e.g.,][]{Singh2011PhPl}. 
Another interesting aspect is that magnetic tension, which is induced by ambipolar diffusion, can be an additional generation mechanism of transverse waves in the chromosphere \citep{Martinez-Sykora_2017Sci}.
In future, the effect of partial ionization should be investigated also in the context of the solar/stellar wind studies.

\subsection{Density fluctuation}
\begin{figure}[ht]
    \begin{center}
    \includegraphics[width=8cm]{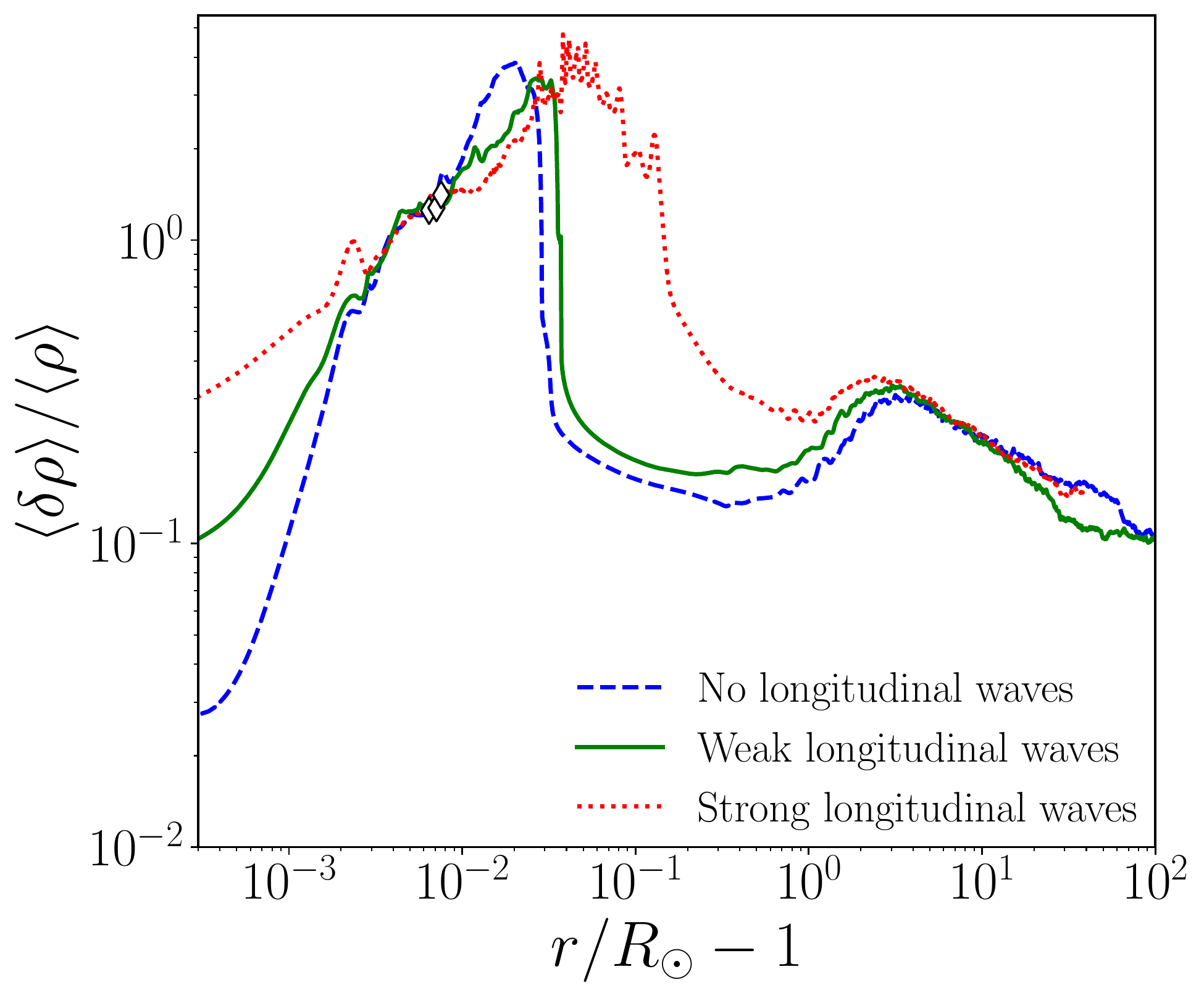}
    \end{center}
    \caption{\label{fig:den_fluc}
    Comparison of the radial profiles of the root-mean-squared density fluctuations normalized by the average density $\langle \delta \rho\rangle / \langle\rho\rangle=\sqrt{\langle\rho^2\rangle-\rho_{\rm ave}^2}/\langle\rho\rangle$ of 
    the three cases, BsV00 (dashed blue), BsV06 (solid green), and BsV18 (dotted red). 
    Diamonds show the location of the top of the chromosphere at $T = 2\times10^4$ K for each case.
    }
\end{figure}

Figure \ref{fig:den_fluc} compares the radial profiles of normalized density fluctuations of different $\langle \delta v_{\parallel,\odot}\rangle$ cases. 
Density fluctuations are
larger near the surface for larger $\langle \delta v_{\parallel,\odot}\rangle$ simply due to the larger injection of the longitudinal perturbations.
However, $\langle \delta \rho\rangle / \langle \rho\rangle$ of the different cases converges to a similar trend in the chromosphere of $r/R_{\odot}-1\gtrsim 3\times 10^{-2}$.
This saturation possibly comes from more efficient longitudinal-to-transverse mode conversion in the chromosphere (Figure \ref{fig:alfven_energy}) in addition to more rapid dissipation of longitudinal wave by shock formation.

Above that, all the presented three cases exhibit a first peak of $\langle \delta \rho\rangle / \langle \rho\rangle$ around  $r/R_{\odot}-1\sim (2-5)\times 10^{-2}$ from the transition region to the low corona.
This peak reflects time-variable spicule activities (see Section \ref{sec:TimeVariability}); 
density fluctuations are excited by the nonlinear mode conversion from transverse waves to longitudinal waves via the gradient of the magnetic pressure associated with the Alfv\'enic waves
\citep{Hollweg_1971_JGR,Kudoh_1999ApJ,Matsumoto_2010_ApJ}.
As explained in Figure \ref{fig:alfven_energy}, the upward Alfv\'enic Poynting flux is larger for larger longitudinal-wave injection at the photosphere even though the same amplitude of transverse fluctuations is excited.
Therefore, taller spicules are generated and the first peak is located at a higher altitude for larger $\langle \delta v_{\parallel,\odot}\rangle$. 

Although the location of the first peak depends on $\langle \delta v_{\parallel,\odot}\rangle$, the radial profiles of $\langle \delta \rho\rangle / \langle \rho \rangle$ converge to a similar level above $r/R_{\odot}\gtrsim 1$ 
almost independently from $\langle \delta v_{\parallel,\odot}\rangle$.
A gentle second peak of $\langle \delta \rho \rangle / \langle \rho \rangle$ is formed around $r/R_{\odot}\sim 5$ by the parametric decay instability of Alfv\'{e}nic waves \citep{Terasawa_1986,Tenerani_2017ApJ,Suzuki_2006_JGRA, Shoda_2018ApJ_PDI,Reville_2018ApJ}.

Since the density fluctuation in the corona and solar wind is observable by remote sensing,
our model could be constrained by the detailed comparison with observation \citep{Miyamoto_2014_ApJ,Hahn_2018_ApJ,Krupar_2020_ApJ}.
For example, it is reported that the relative density fluctuation in the coronal base is as large as $10 \ \%$ or larger \citep{Hahn_2018_ApJ,Krupar_2020_ApJ}, which possibly indicates the non-negligible fraction of compressional waves present in the coronal base.

\begin{figure}[!t]
    \begin{center}
    \includegraphics[width=8cm]{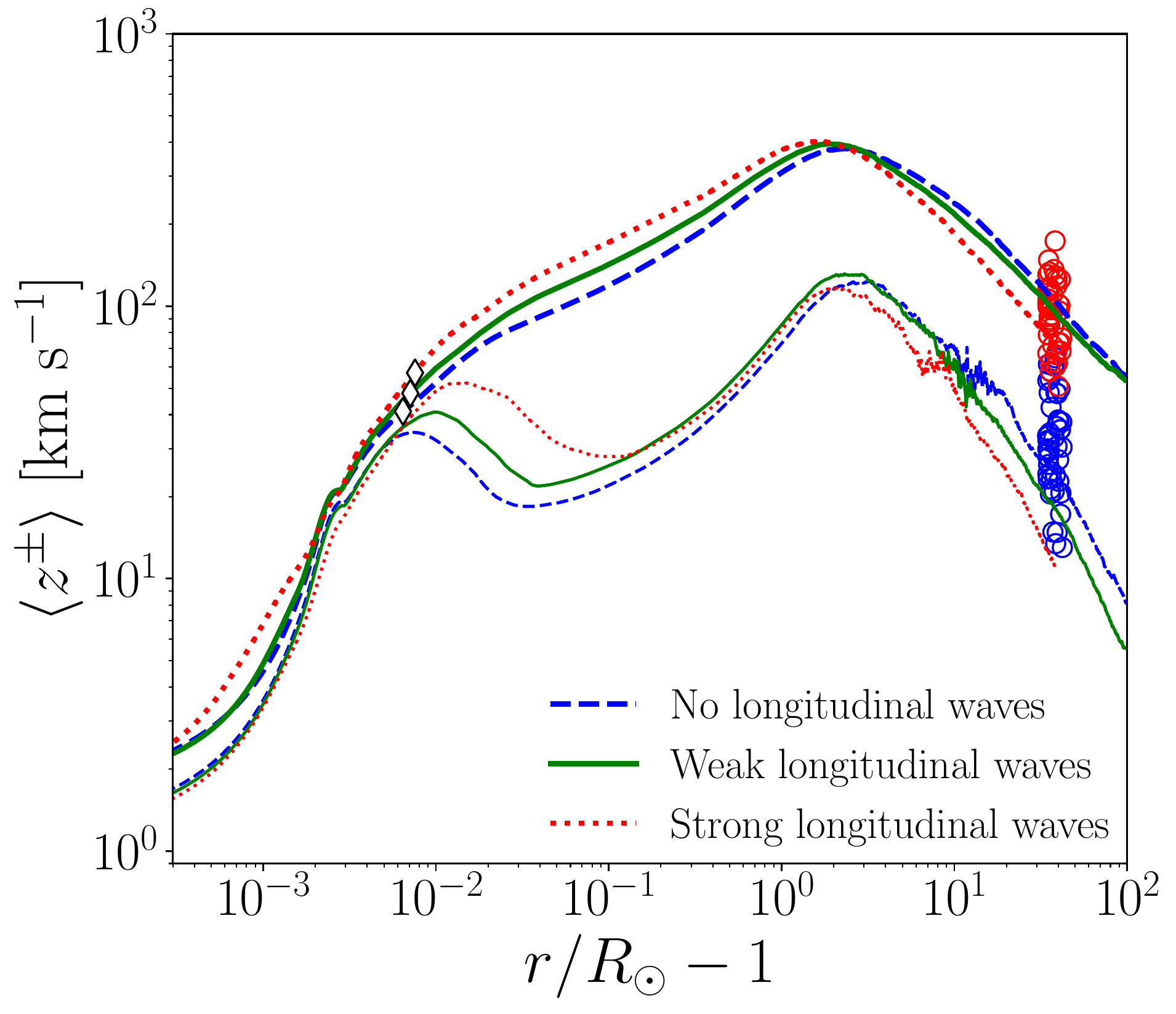}
    \end{center}
    \caption{\label{fig:elssaser}
    Comparison of the radial profiles of the root-mean-squared Els\"asser variables of the outward (thick lines) and inward (thin lines) components. The line types are the same as those in Figure \ref{fig:den_fluc}.
    Red and blue circles respectively represent the observed amplitudes of $z_\perp^+$ and $z_\perp^-$ from PSP \citep{Chen_2020_ApjS}.
    Diamonds show the location of the top of the chromosphere at $T = 2\times10^4$ K for each case.
    }
\end{figure}

\subsection{Alfv\'enicity of the solar wind}
In most cases, the simulated solar wind is fast, in that the termination velocity mostly exceeds $500 {\rm \ km \ s^{-1}}$.
The fast solar wind is known to be Alfv\'enic, that is, the outward Els\"asser variable is much larger than the inward Els\"asser variable in the fast streams.
Here the Alfv\'enic nature of the simulated solar wind is discussed.

Figure \ref{fig:elssaser} compares the numerical results of the time averaged outgoing and incoming Els\"asser variables $\langle z_{+}\rangle$ and $\langle z_{-}\rangle$ to observations at $r\approx 40 R_{\odot}$ by the $\it Parker\ Solar\ Probe$ (hereafter PSP) \citep{Chen_2020_ApjS}. 
The obtained $\langle z_{\pm}\rangle$ from these three cases are consistent with the observed $z_{+}$ and $z_{-}$ that show large scatters.
Both $\langle z_{+}\rangle$ and $\langle z_{-}\rangle$ of our numerical results are larger for larger $\langle \delta v_{\parallel,\odot}\rangle$ in the coronal regions of $10^{-2}<r/R_\odot-1<1$. 
However, larger $\langle z^{\pm}\rangle $ yields larger turbulent loss as shown in Figure \ref{fig:alfven_energy} , which suppresses the increase of $\langle z_{\pm}\rangle$. As a result, almost the same maximum amplitudes of $\langle z_{\pm}\rangle$ are obtained for the different $\langle \delta v_{\parallel,\odot}\rangle$ cases in a self-regulated manner, similarly to the density perturbations (Figure \ref{fig:den_fluc}). In the solar wind region, $r/R_{\odot}\gtrsim 10$, $\langle z_{\pm}\rangle$ is smaller for larger $\langle \delta v_{\parallel,\odot}\rangle$ because the density is higher (Figure \ref{fig:rho}) .

\subsection{Time Variability}
\label{sec:TimeVariability}
\begin{figure*}[!t]
    \centering
    \includegraphics[width=16cm]{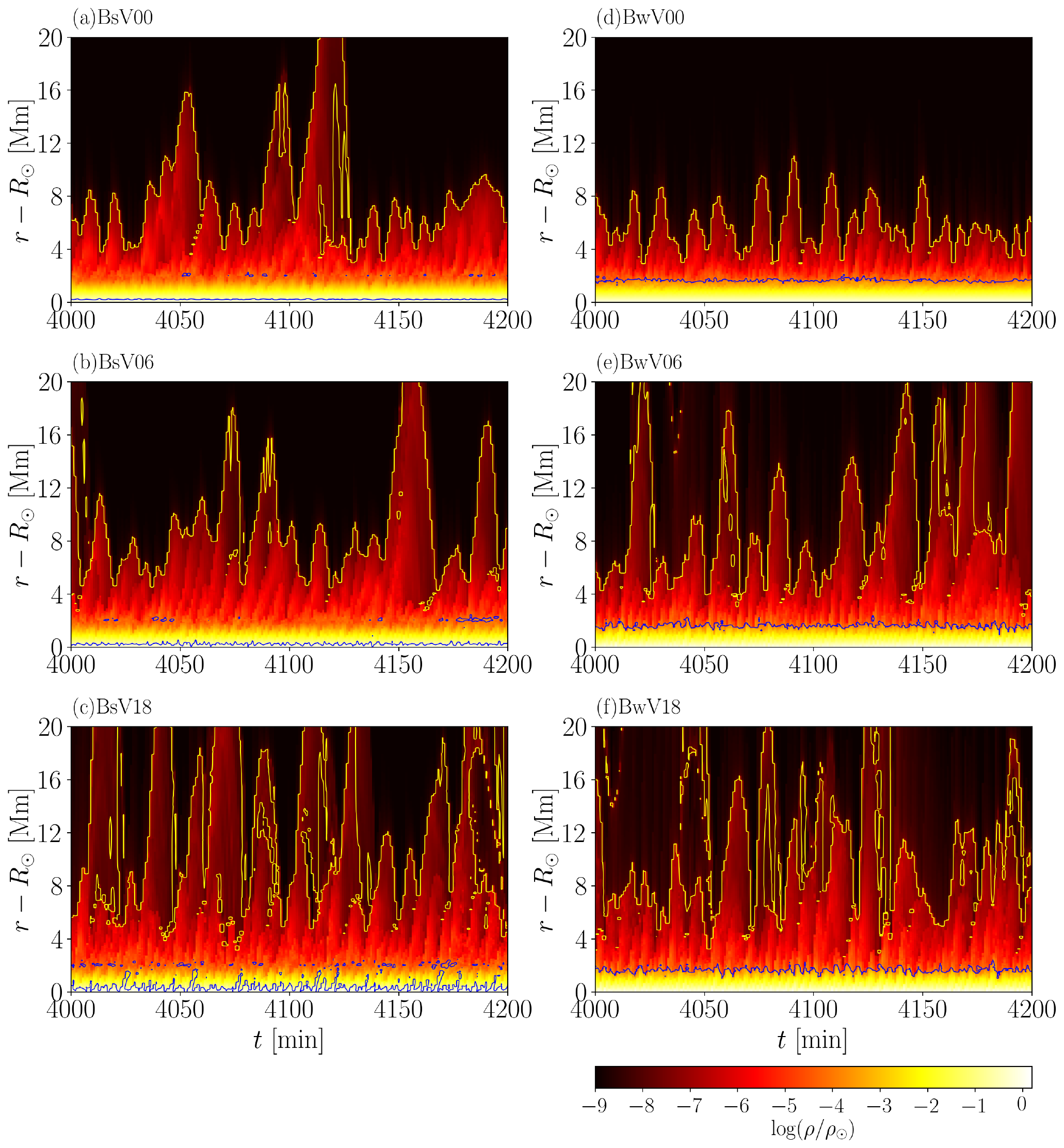}
    \caption{
    \label{fig:r-t_diagram}
    Time-distance diagrams of mass density in the lower atmospheres of BsV00 (a), BsV06 (b), BsV18 (c), BwV00 (d), BwV06 (e) and BwV18 (f).
    The yellow and blue solid lines represent contour lines of 
    $T=2\times 10^4$[K] and $\beta_r=1$, respectively.
    }
\end{figure*}

Figure \ref{fig:r-t_diagram} shows time versus radial distance diagrams of the mass density in the low atmosphere. 
The left and right columns respectively present
the cases with the standard (BsVyy) and weak (BwVyy) magnetic field in the chromosphere. 
The top, middle, and bottom rows 
correspond to $\langle\delta v_{\parallel,\odot}\rangle=$ 0, 0.6, and 1.8 $\rm km \,s^{-1}$, respectively.
The yellow lines represent the positions of $T=2\times 10^4 \,{\rm K}$, which correspond to the bottom of the transition region.

One can see that the transition region moves up and down in all the cases, which should be observed as spicules \citep[e.g.,][]{Beckers_1972ARA&A,De_Pontieu_2007_Science,Shoji_2010_pasj,Okamoto_2011ApJ,Yoshida_2019ApJ,Tei_2020ApJ}. 
The velocity of the upward motions can be derived to be an order of $\sim 10$ km s$^{-1}$ from the slope of yellow lines, which roughly coincides with the sound speed and accordingly the propagation speed of slow MHD waves. 
Namely the gas in the upper chromosphere is lifted up by longitudinal slow-mode waves that are generated from transverse waves through the mode conversion in the upper chromosphere \citep{Hollweg_1982_ApJ,Suematsu_1982SoPh,Matsumoto_2010_ApJ,Sakaue_2021ApJ}.

In the cases without longitudinal perturbations (top panels of Figure \ref{fig:r-t_diagram}),
the stronger field in the chromosphere (top-left) gives the higher and more dynamical transition region because of the larger Alfv\'enic Poynting flux (Figure \ref{fig:alfven_energy2} and Section \ref{sec:dependenceonB}). 
By adding the longitudinal perturbations at the photosphere, the transition region behaves more abruptly and the average height increases.
The comparison of the middle right panel to the top right panel indicates that the activity of the transition region is drastically enhanced in the weak field case by the addition of $\langle \delta v_{\parallel,\odot}\rangle = 0.6$ km s$^{-1}$. 
This is because in the weak field condition transverse waves are generated more effectively from acoustic waves around $\beta_r= 1$ (blue line) in the chromosphere as discussed in Section \ref{sec:dependenceonB}. 

The bottom panels of Figure \ref{fig:r-t_diagram} exhibit a dynamical behavior of transition regions with chromospheric gas being violently uplifted to higher altitudes by the injection of the large-amplitude vertical perturbation with $\langle \delta v_{\parallel,\odot}\rangle = 1.8$ km s$^{-1}$. 
Multiple yellow lines are frequently plotted at single time slices. This indicates that cooler gas with $T<2\times 10^4$ K is transiently distributed above hotter gas with $T>2\times 10^4$ K.

\section{Summary}
\label{sec:summary}
We investigated how the properties of solar winds are 
affected by the $p$-mode like longitudinal perturbation at the photosphere.
We performed 1D simulations from the photosphere to  \shimizu{beyond several tens of solar radii}
for Alfv\'en-wave-driven winds in the wide range of the amplitude of the vertical perturbation of $0.0 {\rm \ km \ s^{-1}} \le \langle\delta v_{\parallel,\odot}\rangle \le 3.0$ km s$^{-1}$ in super-radially open magnetic flux tubes.

The coronal temperature and wind velocity are not significantly affected by the additional input of the longitudinal perturbation (Figure \ref{fig:temp_vr}).  
However, higher coronal density is obtained for larger $\langle\delta v_{\parallel,\odot}\rangle$ (Figure \ref{fig:rho}), and accordingly, the mass-loss rate increases with $\langle\delta v_{\parallel,\odot}\rangle$ by up to $\approx$ 4 times (Figure \ref{fig:vr_mass_loss}) because larger Alfv\'enic Poynting flux enters the corona so as to drive denser outflows 
\shimizu{as a result of more efficient chromospheric evaporation.}
The $p$-mode like vertical oscillation excites acoustic waves, a part of which is converted to the transverse waves by the mode conversion in the chromosphere (Figure \ref{fig:alfven_energy}). 
These transverse waves contribute to the upgoing Alfv\'enic Poynting flux, in addition to the Alfv\'en waves that come from the photosphere.  
This result confirms the observationally inferred link between $p$-mode oscillations and Alfv\'enic waves in the solar corona \citep{Morton_2019_Nature_Astronomy}.  

Cases with larger $\langle\delta v_{\parallel,\odot}\rangle$ exhibit higher time variability and larger density perturbations in the low corona. 
The mass loss rate saturates when $\langle\delta v_{\parallel,\odot}\rangle\gtrsim 2.5$ km s$^{-1}$,
because an increase of $\langle\delta v_{\parallel,\odot}\rangle$ no longer leads to the excitation of transverse waves by the mode conversion but instead is compensated by the radiative loss by the direct shock dissipation of acoustic waves in the chromosphere. 

Simulations with weaker field strength in the low atomosphere show that the magnetic field in the chromosphere controls the mode conversion between longitudinal and transverse modes. 
In the cases that include a region with plasma $\beta\approx 1$ in the middle chromosphere, the mode conversion effectively generates transverse waves there even for a moderate amplitude of $\langle\delta v_{\parallel,\odot}\rangle = 0.6$ km s$^{-1}$.  

We conclude that $p$-mode oscillations at the photosphere play an important role in enhancing Alfv\'enic Poynting flux over the corona of the Sun and solar-type stars.

Numerical simulations in this work were partly carried out on Cray XC50 at Center for Computational Astrophysics, National
Astronomical Observatory of Japan.
M.S. is supported by a Grant-in-Aid for Japan Society for the Promotion of Science (JSPS) Fellows and by the NINS program for cross-disciplinary study (grant Nos. 01321802 and 01311904) on Turbulence, Transport, and Heating Dynamics in Laboratory and Solar/ Astrophysical Plasmas: “SoLaBo-X.”
T.K.S. is supported in part by Grants-in-Aid for Scientific Research from the MEXT/JSPS of Japan, 17H01105 and 21H00033 and by Program for Promoting Research on the Supercomputer Fugaku by the RIKEN Center for Computational Science (Toward a unified view of the universe: from large-scale structures to planets, grant 20351188—PI J. Makino) from the MEXT of Japan.

\newpage

\begin{appendix}
\section{Acoustic wave-driven wind}\label{sec:sound wave wind}

\begin{figure}[htp]
    \begin{center}
    \includegraphics[width=7cm]{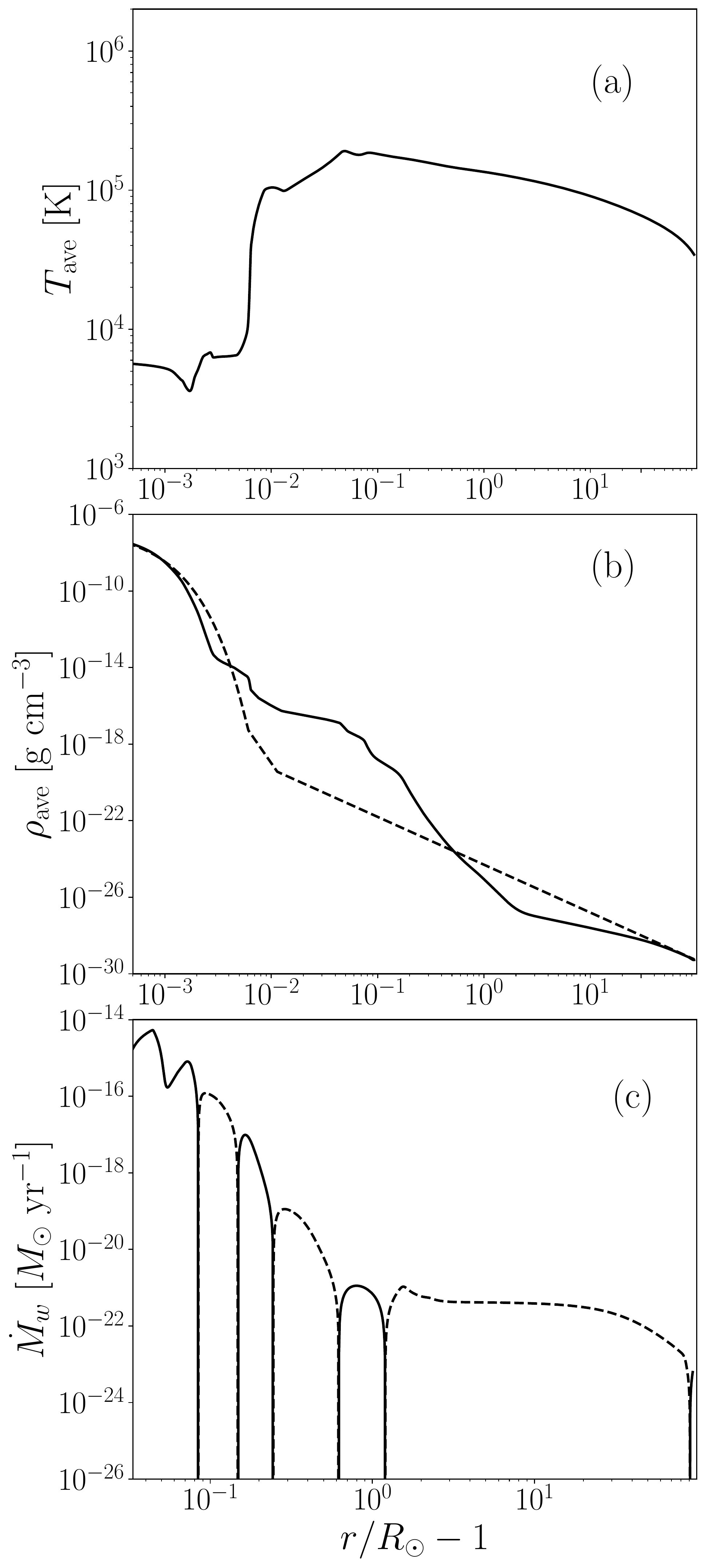}
    \end{center}
    \caption{\label{fig:tate}
    The time-averaged radial profile of B0V06.
    (a): Temperature.
    (b): Density. 
    The black dotted and red solid lines represent the initial profile and the final profile, respectively.
    (c): Mass loss rate.
    Black dotted and solid lines represent negative (inflow) and positive (outflow) values, respectively.
    }
\end{figure}
We examine the properties of the atmosphere of the case with the only longitudinal fluctuation (B0V06) to see if the corona and solar wind are formed solely by acoustic waves. 
In order to avoid the initial infall of material from the upper layer, we set lower initial density ($\rho_{\rm w,0}=10^{-25}$ g cm$^{-3}$ in Eq. \eqref{eq:initdens})
than that of the other cases.

Figure \ref{fig:tate} presents the radial profile of the atmosphere averaged from $t=3500$ min to $t=5000$ min. 
The acoustic waves that travel upward from the photosphere rapidly dissipate at low altitudes, $r/R_{\odot}-1\lesssim 0.1$. 
Although the atmosphere is heated up by wave dissipation of longitudinal waves, 
the temperature of the ``corona'' remains low ($\lesssim 2\times 10^5$ K, Figure \ref{fig:tate} (a))
As a result, the gas in the upper atmosphere does not stream out. 
Instead, it falls down to the surface, which is seen as negative mass loss rate, $\dot{M}_{w}<0$, in $r/R_{\odot}-1 > 1$ (Figure \ref{fig:tate} (c)). The accretion reduces (raises) the density in the outer (inner) region of $r/R_{\odot}-1 <$($>$)$0.5$ from the initial value (Figure \ref{fig:tate} (b)).

The accretion occurs partially because the initial density in the upper region is still higher than the hydrostatic density with $T\lesssim 10^5$ K. 
We note, however, that the initial density is much lower by 6 - 7 orders of magnitude than the observed density in the solar wind. 
This simulation demonstrates that even such low-density gas cannot be driven outward only by the acoustic waves. 
We thus conclude, through a direct numerical demonstration, that the solar coronal heating and the solar wind driving cannot be accomplished only by the sound waves from the photosphere. 
\end{appendix}

\bibliographystyle{aasjournal}
\bibliography{reference.bib}

\begin{thebibliography}{}
\expandafter\ifx\csname natexlab\endcsname\relax\def\natexlab#1{#1}\fi
\providecommand{\url}[1]{\href{#1}{#1}}
\providecommand{\dodoi}[1]{doi:~\href{http://doi.org/#1}{\nolinkurl{#1}}}
\providecommand{\doeprint}[1]{\href{http://ascl.net/#1}{\nolinkurl{http://ascl.net/#1}}}
\providecommand{\doarXiv}[1]{\href{https://arxiv.org/abs/#1}{\nolinkurl{https://arxiv.org/abs/#1}}}

\bibitem[{{Adhikari} {et~al.}(2020){Adhikari}, {Zank}, {Zhao}, {Kasper},
  {Korreck}, {Stevens}, {Case}, {Whittlesey}, {Larson}, {Livi}, \&
  {Klein}}]{Adhikari_2020ApJS}
{Adhikari}, L., {Zank}, G.~P., {Zhao}, L.~L., {et~al.} 2020, \apjs, 246, 38,
  \dodoi{10.3847/1538-4365/ab5852}

\bibitem[{{Alazraki} \& {Couturier}(1971)}]{Alazraki_1971A&A}
{Alazraki}, G., \& {Couturier}, P. 1971, \aap, 13, 380

\bibitem[{{Antolin} {et~al.}(2015){Antolin}, {Okamoto}, {De Pontieu},
  {Uitenbroek}, {Van Doorsselaere}, \& {Yokoyama}}]{Antolin2015_ApJ}
{Antolin}, P., {Okamoto}, T.~J., {De Pontieu}, B., {et~al.} 2015, \apj, 809,
  72, \dodoi{10.1088/0004-637X/809/1/72}

\bibitem[{{Arber} {et~al.}(2016){Arber}, {Brady}, \& {Shelyag}}]{Arber2016ApJ}
{Arber}, T.~D., {Brady}, C.~S., \& {Shelyag}, S. 2016, \apj, 817, 94,
  \dodoi{10.3847/0004-637X/817/2/94}

\bibitem[{{Beckers}(1972)}]{Beckers_1972ARA&A}
{Beckers}, J.~M. 1972, \araa, 10, 73,
  \dodoi{10.1146/annurev.aa.10.090172.000445}

\bibitem[{{Belcher}(1971)}]{Belcher_1971ApJ}
{Belcher}, J.~W. 1971, \apj, 168, 509, \dodoi{10.1086/151105}

\bibitem[{{Biermann}(1951)}]{Biremann_1951ZA}
{Biermann}, L. 1951, \zap, 29, 274

\bibitem[{{Bogdan} \& {Knoelker}(1991)}]{Bogdan_1991ApJ}
{Bogdan}, T.~B., \& {Knoelker}, M. 1991, \apj, 369, 219, \dodoi{10.1086/169753}

\bibitem[{{Bogdan} {et~al.}(1996){Bogdan}, {Knoelker}, {MacGregor}, \&
  {Kim}}]{Bogdan_1996ApJ}
{Bogdan}, T.~J., {Knoelker}, M., {MacGregor}, K.~B., \& {Kim}, E.~J. 1996,
  \apj, 456, 879, \dodoi{10.1086/176704}

\bibitem[{{Bogdan} {et~al.}(2003){Bogdan}, {Carlsson}, {Hansteen}, {McMurry},
  {Rosenthal}, {Johnson}, {Petty-Powell}, {Zita}, {Stein}, {McIntosh}, \&
  {Nordlund}}]{Bogdan_2003ApJ}
{Bogdan}, T.~J., {Carlsson}, M., {Hansteen}, V.~H., {et~al.} 2003, \apj, 599,
  626, \dodoi{10.1086/378512}

\bibitem[{{Cally}(2006)}]{Cally_2006}
{Cally}, P.~S. 2006, Philosophical Transactions of the Royal Society of London
  Series A, 364, 333, \dodoi{10.1098/rsta.2005.1702}

\bibitem[{{Cally} \& {Goossens}(2008)}]{Cally_2008}
{Cally}, P.~S., \& {Goossens}, M. 2008, \solphys, 251, 251,
  \dodoi{10.1007/s11207-007-9086-3}

\bibitem[{{Cally} \& {Hansen}(2011)}]{Cally_2011ApJ}
{Cally}, P.~S., \& {Hansen}, S.~C. 2011, \apj, 738, 119,
  \dodoi{10.1088/0004-637X/738/2/119}

\bibitem[{{Carlsson} \& {Leenaarts}(2012)}]{Carlsson_2012A&A}
{Carlsson}, M., \& {Leenaarts}, J. 2012, \aap, 539, A39,
  \dodoi{10.1051/0004-6361/201118366}

\bibitem[{{Chen} {et~al.}(2020){Chen}, {Bale}, {Bonnell}, {Borovikov}, {Bowen},
  {Burgess}, {Case}, {Chandran}, {de Wit}, {Goetz}, {Harvey}, {Kasper},
  {Klein}, {Korreck}, {Larson}, {Livi}, {MacDowall}, {Malaspina}, {Mallet},
  {McManus}, {Moncuquet}, {Pulupa}, {Stevens}, \&
  {Whittlesey}}]{Chen_2020_ApjS}
{Chen}, C.~H.~K., {Bale}, S.~D., {Bonnell}, J.~W., {et~al.} 2020, \apjs, 246,
  53, \dodoi{10.3847/1538-4365/ab60a3}

\bibitem[{{Chitta} {et~al.}(2012){Chitta}, {van Ballegooijen}, {Rouppe van der
  Voort}, {DeLuca}, \& {Kariyappa}}]{Chitta_2012_ApJ}
{Chitta}, L.~P., {van Ballegooijen}, A., {Rouppe van der Voort}, L., {DeLuca},
  E., \& {Kariyappa}, R. 2012, in American Astronomical Society Meeting
  Abstracts, Vol. 220, American Astronomical Society Meeting Abstracts \#220,
  206.14

\bibitem[{{Cranmer} \& {Saar}(2011)}]{Cranmer_2011_ApJ}
{Cranmer}, S.~R., \& {Saar}, S.~H. 2011, \apj, 741, 54,
  \dodoi{10.1088/0004-637X/741/1/54}

\bibitem[{{Cranmer} {et~al.}(2007){Cranmer}, {van Ballegooijen}, \&
  {Edgar}}]{Cranmer_2007ApJS}
{Cranmer}, S.~R., {van Ballegooijen}, A.~A., \& {Edgar}, R.~J. 2007, \apjs,
  171, 520, \dodoi{10.1086/518001}

\bibitem[{{De Moortel} {et~al.}(2000){De Moortel}, {Hood}, \&
  {Arber}}]{DeMoortel2000A&A}
{De Moortel}, I., {Hood}, A.~W., \& {Arber}, T.~D. 2000, \aap, 354, 334

\bibitem[{{De Pontieu} {et~al.}(2007){De Pontieu}, {McIntosh}, {Carlsson},
  {Hansteen}, {Tarbell}, {Schrijver}, {Title}, {Shine}, {Tsuneta}, {Katsukawa},
  {Ichimoto}, {Suematsu}, {Shimizu}, \& {Nagata}}]{De_Pontieu_2007_Science}
{De Pontieu}, B., {McIntosh}, S.~W., {Carlsson}, M., {et~al.} 2007, Science,
  318, 1574, \dodoi{10.1126/science.1151747}

\bibitem[{Dmitruk {et~al.}(2002)Dmitruk, Matthaeus, Milano, Oughton, Zank, \&
  Mullan}]{Dmitruk_2002}
Dmitruk, P., Matthaeus, W.~H., Milano, L.~J., {et~al.} 2002, The Astrophysical
  Journal, 575, 571, \dodoi{10.1086/341188}

\bibitem[{{Elsasser}(1950)}]{Elsasser_PhRv_1950}
{Elsasser}, W.~M. 1950, Physical Review, 79, 183,
  \dodoi{10.1103/PhysRev.79.183}

\bibitem[{{Farahani} {et~al.}(2021){Farahani}, {Hejazi}, \&
  {Boroomand}}]{Vasheghani_Farahani_2021ApJ}
{Farahani}, S.~V., {Hejazi}, S.~M., \& {Boroomand}, M.~R. 2021, \apj, 906, 70,
  \dodoi{10.3847/1538-4357/abca8c}

\bibitem[{{Felipe} {et~al.}(2010){Felipe}, {Khomenko}, {Collados}, \&
  {Beck}}]{Felipe_2010}
{Felipe}, T., {Khomenko}, E., {Collados}, M., \& {Beck}, C. 2010, \apj, 722,
  131, \dodoi{10.1088/0004-637X/722/1/131}

\bibitem[{{Fisk}(2003)}]{Fisk_2003JGRA}
{Fisk}, L.~A. 2003, Journal of Geophysical Research (Space Physics), 108, 1157,
  \dodoi{10.1029/2002JA009284}

\bibitem[{{Fludra} {et~al.}(1999){Fludra}, {Del Zanna}, \&
  {Bromage}}]{Fludra_1999SSRv}
{Fludra}, A., {Del Zanna}, G., \& {Bromage}, B.~J.~I. 1999, \ssr, 87, 185,
  \dodoi{10.1023/A:1005127930584}

\bibitem[{{Geiss} {et~al.}(1995){Geiss}, {Gloeckler}, \& {von
  Steiger}}]{Geiss_1995SSRev}
{Geiss}, J., {Gloeckler}, G., \& {von Steiger}, R. 1995, \ssr, 72, 49,
  \dodoi{10.1007/BF00768753}

\bibitem[{{Goldreich} \& {Sridhar}(1995)}]{Goldreich_1995ApJ}
{Goldreich}, P., \& {Sridhar}, S. 1995, \apj, 438, 763, \dodoi{10.1086/175121}

\bibitem[{{Goodman} \& {Judge}(2012)}]{Googman_2012ApJ}
{Goodman}, M.~L., \& {Judge}, P.~G. 2012, \apj, 751, 75,
  \dodoi{10.1088/0004-637X/751/1/75}

\bibitem[{{G{\"u}del} {et~al.}(2014){G{\"u}del}, {Dvorak}, {Erkaev}, {Kasting},
  {Khodachenko}, {Lammer}, {Pilat-Lohinger}, {Rauer}, {Ribas}, \&
  {Wood}}]{Gudel_2014_proceedings}
{G{\"u}del}, M., {Dvorak}, R., {Erkaev}, N., {et~al.} 2014, in Protostars and
  Planets VI, ed. H.~{Beuther}, R.~S. {Klessen}, C.~P. {Dullemond}, \&
  T.~{Henning}, 883, \dodoi{10.2458/azu_uapress_9780816531240-ch038}

\bibitem[{{Gudiksen} \& {Nordlund}(2005)}]{Gudiksen_Nordlund_2005_Apj}
{Gudiksen}, B.~V., \& {Nordlund}, {\r{A}}. 2005, \apj, 618, 1020,
  \dodoi{10.1086/426063}

\bibitem[{{Hahn} {et~al.}(2018){Hahn}, {D'Huys}, \& {Savin}}]{Hahn_2018_ApJ}
{Hahn}, M., {D'Huys}, E., \& {Savin}, D.~W. 2018, \apj, 860, 34,
  \dodoi{10.3847/1538-4357/aac0f3}

\bibitem[{{Hansteen} {et~al.}(2015){Hansteen}, {Guerreiro}, {De Pontieu}, \&
  {Carlsson}}]{Hansteen2015ApJ}
{Hansteen}, V., {Guerreiro}, N., {De Pontieu}, B., \& {Carlsson}, M. 2015,
  \apj, 811, 106, \dodoi{10.1088/0004-637X/811/2/106}

\bibitem[{{Hansteen} \& {Leer}(1995)}]{Hansteen_1995_JGR}
{Hansteen}, V.~H., \& {Leer}, E. 1995, \jgr, 100, 21577,
  \dodoi{10.1029/95JA02300}

\bibitem[{{Hasan} \& {van Ballegooijen}(2008)}]{Hasan_2008ApJ}
{Hasan}, S.~S., \& {van Ballegooijen}, A.~A. 2008, \apj, 680, 1542,
  \dodoi{10.1086/587773}

\bibitem[{{Heyvaerts} \& {Priest}(1983)}]{Heyvaerts_1983_AA}
{Heyvaerts}, J., \& {Priest}, E.~R. 1983, \aap, 117, 220

\bibitem[{{Hollweg}(1971)}]{Hollweg_1971_JGR}
{Hollweg}, J.~V. 1971, \jgr, 76, 5155, \dodoi{10.1029/JA076i022p05155}

\bibitem[{{Hollweg}(1982)}]{Hollweg_1982_ApJ}
---. 1982, \apj, 257, 345, \dodoi{10.1086/159993}

\bibitem[{{Hossain} {et~al.}(1995){Hossain}, {Gray}, {Pontius}, {Matthaeus}, \&
  {Oughton}}]{Hossain_1995PhFl}
{Hossain}, M., {Gray}, P.~C., {Pontius}, Duane~H., J., {Matthaeus}, W.~H., \&
  {Oughton}, S. 1995, Physics of Fluids, 7, 2886, \dodoi{10.1063/1.868665}

\bibitem[{{Howes} \& {Nielson}(2013)}]{Howes_2013PhPl}
{Howes}, G.~G., \& {Nielson}, K.~D. 2013, Physics of Plasmas, 20, 072302,
  \dodoi{10.1063/1.4812805}

\bibitem[{{Iijima} \& {Yokoyama}(2017)}]{Iijima_2017ApJ}
{Iijima}, H., \& {Yokoyama}, T. 2017, \apj, 848, 38,
  \dodoi{10.3847/1538-4357/aa8ad1}

\bibitem[{{Ionson}(1978)}]{Ionson_1978_ApJ}
{Ionson}, J.~A. 1978, \apj, 226, 650, \dodoi{10.1086/156648}

\bibitem[{{Ito} {et~al.}(2010){Ito}, {Tsuneta}, {Shiota}, {Tokumaru}, \&
  {Fujiki}}]{Ito_2010ApJ}
{Ito}, H., {Tsuneta}, S., {Shiota}, D., {Tokumaru}, M., \& {Fujiki}, K. 2010,
  \apj, 719, 131, \dodoi{10.1088/0004-637X/719/1/131}

\bibitem[{{Khodachenko} {et~al.}(2004){Khodachenko}, {Arber}, {Rucker}, \&
  {Hanslmeier}}]{Khodachenko2004A&A}
{Khodachenko}, M.~L., {Arber}, T.~D., {Rucker}, H.~O., \& {Hanslmeier}, A.
  2004, \aap, 422, 1073, \dodoi{10.1051/0004-6361:20034207}

\bibitem[{{Khomenko} \& {Collados}(2012)}]{Khomenko_2012ApJ}
{Khomenko}, E., \& {Collados}, M. 2012, \apj, 747, 87,
  \dodoi{10.1088/0004-637X/747/2/87}

\bibitem[{{Klimchuk}(2006)}]{Klimchuk_2006_SolPhys}
{Klimchuk}, J.~A. 2006, \solphys, 234, 41, \dodoi{10.1007/s11207-006-0055-z}

\bibitem[{{Kohl} {et~al.}(2006){Kohl}, {Noci}, {Cranmer}, \&
  {Raymond}}]{Kohl_2006A&ARv}
{Kohl}, J.~L., {Noci}, G., {Cranmer}, S.~R., \& {Raymond}, J.~C. 2006, \aapr,
  13, 31, \dodoi{10.1007/s00159-005-0026-7}

\bibitem[{{Kopp} \& {Holzer}(1976)}]{Kopp_1976_SolPhys}
{Kopp}, R.~A., \& {Holzer}, T.~E. 1976, \solphys, 49, 43,
  \dodoi{10.1007/BF00221484}

\bibitem[{{Krieger} {et~al.}(1973){Krieger}, {Timothy}, \&
  {Roelof}}]{Krieger_1973SoPh}
{Krieger}, A.~S., {Timothy}, A.~F., \& {Roelof}, E.~C. 1973, \solphys, 29, 505,
  \dodoi{10.1007/BF00150828}

\bibitem[{{Krupar} {et~al.}(2020){Krupar}, {Szabo}, {Maksimovic}, {Kruparova},
  {Kontar}, {Balmaceda}, {Bonnin}, {Bale}, {Pulupa}, {Malaspina}, {Bonnell},
  {Harvey}, {Goetz}, {Dudok de Wit}, {MacDowall}, {Kasper}, {Case}, {Korreck},
  {Larson}, {Livi}, {Stevens}, {Whittlesey}, \& {Hegedus}}]{Krupar_2020_ApJ}
{Krupar}, V., {Szabo}, A., {Maksimovic}, M., {et~al.} 2020, \apjs, 246, 57,
  \dodoi{10.3847/1538-4365/ab65bd}

\bibitem[{{Kudoh} \& {Shibata}(1999)}]{Kudoh_1999ApJ}
{Kudoh}, T., \& {Shibata}, K. 1999, \apj, 514, 493, \dodoi{10.1086/306930}

\bibitem[{{Lamers} \& {Cassinelli}(1999)}]{Lamers_1999book}
{Lamers}, H. J.~G.~L.~M., \& {Cassinelli}, J.~P. 1999, {Introduction to Stellar
  Winds}

\bibitem[{{Lazarian}(2016)}]{Lazarian_2016}
{Lazarian}, A. 2016, \apj, 833, 131, \dodoi{10.3847/1538-4357/833/2/131}

\bibitem[{{Lighthill}(1952)}]{Lighthill_1952RSPSA}
{Lighthill}, M.~J. 1952, Proceedings of the Royal Society of London Series A,
  211, 564, \dodoi{10.1098/rspa.1952.0060}

\bibitem[{{Lionello} {et~al.}(2014){Lionello}, {Velli}, {Downs}, {Linker},
  {Miki{\'c}}, \& {Verdini}}]{Lionello_2014_ApJ}
{Lionello}, R., {Velli}, M., {Downs}, C., {et~al.} 2014, \apj, 784, 120,
  \dodoi{10.1088/0004-637X/784/2/120}

\bibitem[{{Magyar} {et~al.}(2017){Magyar}, {Van Doorsselaere}, \&
  {Goossens}}]{Magyar_2017NatSR}
{Magyar}, N., {Van Doorsselaere}, T., \& {Goossens}, M. 2017, Scientific
  Reports, 7, 14820, \dodoi{10.1038/s41598-017-13660-1}

\bibitem[{{Mart{\'\i}nez-Sykora} {et~al.}(2017){Mart{\'\i}nez-Sykora}, {De
  Pontieu}, {Hansteen}, {Rouppe van der Voort}, {Carlsson}, \&
  {Pereira}}]{Martinez-Sykora_2017Sci}
{Mart{\'\i}nez-Sykora}, J., {De Pontieu}, B., {Hansteen}, V.~H., {et~al.} 2017,
  Science, 356, 1269, \dodoi{10.1126/science.aah5412}

\bibitem[{{Matsumoto}(2021)}]{Matsumoto_2021_MNRAS}
{Matsumoto}, T. 2021, \mnras, 500, 4779, \dodoi{10.1093/mnras/staa3533}

\bibitem[{{Matsumoto} \& {Kitai}(2010)}]{Matsumoto_2010ApJ}
{Matsumoto}, T., \& {Kitai}, R. 2010, \apjl, 716, L19,
  \dodoi{10.1088/2041-8205/716/1/L19}

\bibitem[{{Matsumoto} \& {Shibata}(2010)}]{Matsumoto_2010_ApJ}
{Matsumoto}, T., \& {Shibata}, K. 2010, \apj, 710, 1857,
  \dodoi{10.1088/0004-637X/710/2/1857}

\bibitem[{{Matsumoto} \& {Suzuki}(2012)}]{Matsumoto_2012ApJ}
{Matsumoto}, T., \& {Suzuki}, T.~K. 2012, \apj, 749, 8,
  \dodoi{10.1088/0004-637X/749/1/8}

\bibitem[{{Matsumoto} \& {Suzuki}(2014)}]{Matsumoto_2014MNRAS}
---. 2014, \mnras, 440, 971, \dodoi{10.1093/mnras/stu310}

\bibitem[{{Matthaeus} {et~al.}(1999){Matthaeus}, {Zank}, {Oughton}, {Mullan},
  \& {Dmitruk}}]{Matthaeus_1999_ApJ}
{Matthaeus}, W.~H., {Zank}, G.~P., {Oughton}, S., {Mullan}, D.~J., \&
  {Dmitruk}, P. 1999, \apjl, 523, L93, \dodoi{10.1086/312259}

\bibitem[{{McIntosh} {et~al.}(2011){McIntosh}, {de Pontieu}, {Carlsson},
  {Hansteen}, {Boerner}, \& {Goossens}}]{McIntosh_2011}
{McIntosh}, S.~W., {de Pontieu}, B., {Carlsson}, M., {et~al.} 2011, \nat, 475,
  477, \dodoi{10.1038/nature10235}

\bibitem[{{Miyamoto} {et~al.}(2014){Miyamoto}, {Imamura}, {Tokumaru}, {Ando},
  {Isobe}, {Asai}, {Shiota}, {Toda}, {H{\"a}usler}, {P{\"a}tzold}, {Nabatov},
  \& {Nakamura}}]{Miyamoto_2014_ApJ}
{Miyamoto}, M., {Imamura}, T., {Tokumaru}, M., {et~al.} 2014, \apj, 797, 51,
  \dodoi{10.1088/0004-637X/797/1/51}

\bibitem[{{Morton} {et~al.}(2019){Morton}, {Weberg}, \&
  {McLaughlin}}]{Morton_2019_Nature_Astronomy}
{Morton}, R.~J., {Weberg}, M.~J., \& {McLaughlin}, J.~A. 2019, Nature
  Astronomy, 3, 223, \dodoi{10.1038/s41550-018-0668-9}

\bibitem[{{Narukage} {et~al.}(2011){Narukage}, {Sakao}, {Kano}, {Hara},
  {Shimojo}, {Bando}, {Urayama}, {Deluca}, {Golub}, {Weber}, {Grigis},
  {Cirtain}, \& {Tsuneta}}]{Narukage_2011SoPh}
{Narukage}, N., {Sakao}, T., {Kano}, R., {et~al.} 2011, \solphys, 269, 169,
  \dodoi{10.1007/s11207-010-9685-2}

\bibitem[{{Neugebauer} \& {Snyder}(1966)}]{Neugebauer_1966JGR}
{Neugebauer}, M., \& {Snyder}, C.~W. 1966, \jgr, 71, 4469,
  \dodoi{10.1029/JZ071i019p04469}

\bibitem[{{Nishizuka} {et~al.}(2008){Nishizuka}, {Shimizu}, {Nakamura},
  {Otsuji}, {Okamoto}, {Katsukawa}, \& {Shibata}}]{Nishuzuka_2008ApJ}
{Nishizuka}, N., {Shimizu}, M., {Nakamura}, T., {et~al.} 2008, \apjl, 683, L83,
  \dodoi{10.1086/591445}

\bibitem[{{Oba} {et~al.}(2017{\natexlab{a}}){Oba}, {Riethm{\"u}ller},
  {Solanki}, {Iida}, {Quintero Noda}, \& {Shimizu}}]{Oba2017ApJ}
{Oba}, T., {Riethm{\"u}ller}, T.~L., {Solanki}, S.~K., {et~al.}
  2017{\natexlab{a}}, \apj, 849, 7, \dodoi{10.3847/1538-4357/aa8e44}

\bibitem[{{Oba} {et~al.}(2017{\natexlab{b}}){Oba}, {Riethm{\"u}ller},
  {Solanki}, {Iida}, {Quintero Noda}, \& {Shimizu}}]{Oba_2017ApJ}
---. 2017{\natexlab{b}}, \apj, 849, 7, \dodoi{10.3847/1538-4357/aa8e44}

\bibitem[{{Okamoto} \& {De Pontieu}(2011)}]{Okamoto_2011ApJ}
{Okamoto}, T.~J., \& {De Pontieu}, B. 2011, \apjl, 736, L24,
  \dodoi{10.1088/2041-8205/736/2/L24}

\bibitem[{{Okamoto} {et~al.}(2007){Okamoto}, {Tsuneta}, {Berger}, {Ichimoto},
  {Katsukawa}, {Lites}, {Nagata}, {Shibata}, {Shimizu}, {Shine}, {Suematsu},
  {Tarbell}, \& {Title}}]{Okamoto_2007Sci}
{Okamoto}, T.~J., {Tsuneta}, S., {Berger}, T.~E., {et~al.} 2007, Science, 318,
  1577, \dodoi{10.1126/science.1145447}

\bibitem[{{Parker}(1958)}]{Parker_1958ApJ}
{Parker}, E.~N. 1958, \apj, 128, 664, \dodoi{10.1086/146579}

\bibitem[{{Perez} \& {Chandran}(2013)}]{Perez_2013ApJ}
{Perez}, J.~C., \& {Chandran}, B. D.~G. 2013, \apj, 776, 124,
  \dodoi{10.1088/0004-637X/776/2/124}

\bibitem[{{Priest}(2014)}]{Priest_2014masu.book}
{Priest}, E. 2014, {Magnetohydrodynamics of the Sun},
  \dodoi{10.1017/CBO9781139020732}

\bibitem[{{R{\'e}ville} {et~al.}(2018){R{\'e}ville}, {Tenerani}, \&
  {Velli}}]{Reville_2018ApJ}
{R{\'e}ville}, V., {Tenerani}, A., \& {Velli}, M. 2018, \apj, 866, 38,
  \dodoi{10.3847/1538-4357/aadb8f}

\bibitem[{{Rosner} {et~al.}(1978){Rosner}, {Tucker}, \& {Vaiana}}]{RTV_1978ApJ}
{Rosner}, R., {Tucker}, W.~H., \& {Vaiana}, G.~S. 1978, \apj, 220, 643,
  \dodoi{10.1086/155949}

\bibitem[{{Saito} {et~al.}(1970){Saito}, {Makita}, {Nishi}, \&
  {Hata}}]{Saito_1970AnTok}
{Saito}, K., {Makita}, M., {Nishi}, K., \& {Hata}, S. 1970, Annals of the Tokyo
  Astronomical Observatory, 12, 51

\bibitem[{{Sakaue} \& {Shibata}(2020)}]{Sakaue_2020ApJ}
{Sakaue}, T., \& {Shibata}, K. 2020, \apj, 900, 120,
  \dodoi{10.3847/1538-4357/ababa0}

\bibitem[{{Sakaue} \& {Shibata}(2021)}]{Sakaue_2021ApJ}
---. 2021, \apj, 919, 29, \dodoi{10.3847/1538-4357/ac0e34}

\bibitem[{{Schunker} \& {Cally}(2006)}]{Schunker_2006}
{Schunker}, H., \& {Cally}, P.~S. 2006, \mnras, 372, 551,
  \dodoi{10.1111/j.1365-2966.2006.10855.x}

\bibitem[{{Shiota} {et~al.}(2017){Shiota}, {Zank}, {Adhikari}, {Hunana},
  {Telloni}, \& {Bruno}}]{Shiota_2017ApJ}
{Shiota}, D., {Zank}, G.~P., {Adhikari}, L., {et~al.} 2017, \apj, 837, 75,
  \dodoi{10.3847/1538-4357/aa60bc}

\bibitem[{{Shoda} {et~al.}(2019){Shoda}, {Suzuki}, {Asgari-Targhi}, \&
  {Yokoyama}}]{Shoda_2019ApJ}
{Shoda}, M., {Suzuki}, T.~K., {Asgari-Targhi}, M., \& {Yokoyama}, T. 2019,
  \apjl, 880, L2, \dodoi{10.3847/2041-8213/ab2b45}

\bibitem[{{Shoda} \& {Takasao}(2021)}]{Shoda_Takasao_2021arXiv}
{Shoda}, M., \& {Takasao}, S. 2021, arXiv e-prints, arXiv:2106.08915.
\newblock \doarXiv{2106.08915}

\bibitem[{{Shoda} {et~al.}(2018{\natexlab{a}}){Shoda}, {Yokoyama}, \&
  {Suzuki}}]{Shoda_2018_ApJ_a_self-consistent_model}
{Shoda}, M., {Yokoyama}, T., \& {Suzuki}, T.~K. 2018{\natexlab{a}}, \apj, 853,
  190, \dodoi{10.3847/1538-4357/aaa3e1}

\bibitem[{{Shoda} {et~al.}(2018{\natexlab{b}}){Shoda}, {Yokoyama}, \&
  {Suzuki}}]{Shoda_2018ApJ_PDI}
---. 2018{\natexlab{b}}, \apj, 860, 17, \dodoi{10.3847/1538-4357/aac218}

\bibitem[{{Shoda} {et~al.}(2020){Shoda}, {Suzuki}, {Matt}, {Cranmer},
  {Vidotto}, {Strugarek}, {See}, {R{\'e}ville}, {Finley}, \&
  {Brun}}]{Shoda_2020_ApJ}
{Shoda}, M., {Suzuki}, T.~K., {Matt}, S.~P., {et~al.} 2020, \apj, 896, 123,
  \dodoi{10.3847/1538-4357/ab94bf}

\bibitem[{Shoji {et~al.}(2010)Shoji, Nishikawa, Kitai, \&
  UeNo}]{Shoji_2010_pasj}
Shoji, M., Nishikawa, T., Kitai, R., \& UeNo, S. 2010, Publications of the
  Astronomical Society of Japan, 62, 927, \dodoi{10.1093/pasj/62.4.927}

\bibitem[{{Singh} {et~al.}(2011){Singh}, {Shibata}, {Nishizuka}, \&
  {Isobe}}]{Singh2011PhPl}
{Singh}, K.~A.~P., {Shibata}, K., {Nishizuka}, N., \& {Isobe}, H. 2011, Physics
  of Plasmas, 18, 111210, \dodoi{10.1063/1.3655444}

\bibitem[{{Spitzer} \& {H{\"a}rm}(1953)}]{Spitzer_1953_PhRv}
{Spitzer}, L., \& {H{\"a}rm}, R. 1953, Physical Review, 89, 977,
  \dodoi{10.1103/PhysRev.89.977}

\bibitem[{{Spruit} \& {Bogdan}(1992)}]{Spruit_1992_ApJ}
{Spruit}, H.~C., \& {Bogdan}, T.~J. 1992, \apjl, 391, L109,
  \dodoi{10.1086/186409}

\bibitem[{{Srivastava} {et~al.}(2017){Srivastava}, {Shetye}, {Murawski},
  {Doyle}, {Stangalini}, {Scullion}, {Ray}, {W{\'o}jcik}, \&
  {Dwivedi}}]{Srivastava_2017NatSR}
{Srivastava}, A.~K., {Shetye}, J., {Murawski}, K., {et~al.} 2017, Scientific
  Reports, 7, 43147, \dodoi{10.1038/srep43147}

\bibitem[{{Stein}(1967)}]{Stein_1967SoPh}
{Stein}, R.~F. 1967, \solphys, 2, 385, \dodoi{10.1007/BF00146490}

\bibitem[{{Stein} \& {Schwartz}(1972)}]{Stein_1972ApJ}
{Stein}, R.~F., \& {Schwartz}, R.~A. 1972, \apj, 177, 807,
  \dodoi{10.1086/151757}

\bibitem[{{Stepien}(1988)}]{Stepien1988ApJ}
{Stepien}, K. 1988, \apj, 335, 892, \dodoi{10.1086/166975}

\bibitem[{{Suematsu} {et~al.}(1982){Suematsu}, {Shibata}, {Neshikawa}, \&
  {Kitai}}]{Suematsu_1982SoPh}
{Suematsu}, Y., {Shibata}, K., {Neshikawa}, T., \& {Kitai}, R. 1982, \solphys,
  75, 99, \dodoi{10.1007/BF00153464}

\bibitem[{{Suzuki}(2002)}]{Suzuki_2002ApJ}
{Suzuki}, T.~K. 2002, \apj, 578, 598, \dodoi{10.1086/342347}

\bibitem[{{Suzuki} {et~al.}(2013){Suzuki}, {Imada}, {Kataoka}, {Kato},
  {Matsumoto}, {Miyahara}, \& {Tsuneta}}]{Suzuki_2013_PASJ}
{Suzuki}, T.~K., {Imada}, S., {Kataoka}, R., {et~al.} 2013, \pasj, 65, 98,
  \dodoi{10.1093/pasj/65.5.98}

\bibitem[{{Suzuki} \& {Inutsuka}(2005)}]{Suzuki_2005_ApJ}
{Suzuki}, T.~K., \& {Inutsuka}, S.-i. 2005, \apjl, 632, L49,
  \dodoi{10.1086/497536}

\bibitem[{{Suzuki} \& {Inutsuka}(2006)}]{Suzuki_2006_JGRA}
{Suzuki}, T.~K., \& {Inutsuka}, S.-I. 2006, Journal of Geophysical Research
  (Space Physics), 111, A06101, \dodoi{10.1029/2005JA011502}

\bibitem[{{Tei} {et~al.}(2020){Tei}, {Gun{\'a}r}, {Heinzel}, {Okamoto},
  {{\v{S}}t{\v{e}}p{\'a}n}, {Jej{\v{c}}i{\v{c}}}, \& {Shibata}}]{Tei_2020ApJ}
{Tei}, A., {Gun{\'a}r}, S., {Heinzel}, P., {et~al.} 2020, \apj, 888, 42,
  \dodoi{10.3847/1538-4357/ab5db1}

\bibitem[{{Tenerani} {et~al.}(2017){Tenerani}, {Velli}, \&
  {Hellinger}}]{Tenerani_2017ApJ}
{Tenerani}, A., {Velli}, M., \& {Hellinger}, P. 2017, \apj, 851, 99,
  \dodoi{10.3847/1538-4357/aa9bef}

\bibitem[{{Terasawa} {et~al.}(1986){Terasawa}, {Hoshino}, {Sakai}, \&
  {Hada}}]{Terasawa_1986}
{Terasawa}, T., {Hoshino}, M., {Sakai}, J.~I., \& {Hada}, T. 1986, \jgr, 91,
  4171, \dodoi{10.1029/JA091iA04p04171}

\bibitem[{{Teriaca} {et~al.}(2003){Teriaca}, {Poletto}, {Romoli}, \&
  {Biesecker}}]{Teriaca_2003AIPC}
{Teriaca}, L., {Poletto}, G., {Romoli}, M., \& {Biesecker}, D. 2003, in
  American Institute of Physics Conference Series, Vol. 679, Solar Wind Ten,
  ed. M.~{Velli}, R.~{Bruno}, F.~{Malara}, \& B.~{Bucci}, 327--330,
  \dodoi{10.1063/1.1618605}

\bibitem[{{Thurgood} {et~al.}(2014){Thurgood}, {Morton}, \&
  {McLaughlin}}]{Thurgood_2014ApJ}
{Thurgood}, J.~O., {Morton}, R.~J., \& {McLaughlin}, J.~A. 2014, \apjl, 790,
  L2, \dodoi{10.1088/2041-8205/790/1/L2}

\bibitem[{{Tsuneta} {et~al.}(2008){Tsuneta}, {Ichimoto}, {Katsukawa}, {Lites},
  {Matsuzaki}, {Nagata}, {Orozco Su{\'a}rez}, {Shimizu}, {Shimojo}, {Shine},
  {Suematsu}, {Suzuki}, {Tarbell}, \& {Title}}]{Tsuneta_2008ApJ}
{Tsuneta}, S., {Ichimoto}, K., {Katsukawa}, Y., {et~al.} 2008, \apj, 688, 1374,
  \dodoi{10.1086/592226}

\bibitem[{{van Ballegooijen} \&
  {Asgari-Targhi}(2016)}]{van_Ballegooijen_2016ApJ}
{van Ballegooijen}, A.~A., \& {Asgari-Targhi}, M. 2016, \apj, 821, 106,
  \dodoi{10.3847/0004-637X/821/2/106}

\bibitem[{van Ballegooijen \& Asgari-Targhi(2017)}]{van_Ballegooijen_2017}
van Ballegooijen, A.~A., \& Asgari-Targhi, M. 2017, The Astrophysical Journal,
  835, 10, \dodoi{10.3847/1538-4357/835/1/10}

\bibitem[{{Van Doorsselaere} {et~al.}(2004){Van Doorsselaere}, {Andries},
  {Poedts}, \& {Goossens}}]{VanDoorsselaere2004}
{Van Doorsselaere}, T., {Andries}, J., {Poedts}, S., \& {Goossens}, M. 2004,
  \apj, 606, 1223, \dodoi{10.1086/383191}

\bibitem[{{Vasheghani Farahani} {et~al.}(2012){Vasheghani Farahani},
  {Nakariakov}, {Verwichte}, \& {Van
  Doorsselaere}}]{Vasheghani_Farahani_2012A&A}
{Vasheghani Farahani}, S., {Nakariakov}, V.~M., {Verwichte}, E., \& {Van
  Doorsselaere}, T. 2012, \aap, 544, A127, \dodoi{10.1051/0004-6361/201219569}

\bibitem[{Velli {et~al.}(1989)Velli, Grappin, \&
  Mangeney}]{Velli_1989_PhysRevLett}
Velli, M., Grappin, R., \& Mangeney, A. 1989, Phys. Rev. Lett., 63, 1807,
  \dodoi{10.1103/PhysRevLett.63.1807}

\bibitem[{{Verdini} {et~al.}(2010){Verdini}, {Velli}, {Matthaeus}, {Oughton},
  \& {Dmitruk}}]{Verdini_2010_ApJ}
{Verdini}, A., {Velli}, M., {Matthaeus}, W.~H., {Oughton}, S., \& {Dmitruk}, P.
  2010, \apjl, 708, L116, \dodoi{10.1088/2041-8205/708/2/L116}

\bibitem[{{Vidotto}(2021)}]{Vidotto2021LRSP}
{Vidotto}, A.~A. 2021, Living Reviews in Solar Physics, 18, 3,
  \dodoi{10.1007/s41116-021-00029-w}

\bibitem[{{von Steiger} {et~al.}(2000){von Steiger}, {Schwadron}, {Fisk},
  {Geiss}, {Gloeckler}, {Hefti}, {Wilken}, {Wimmer-Schweingruber}, \&
  {Zurbuchen}}]{von_Steiger_2000JGR}
{von Steiger}, R., {Schwadron}, N.~A., {Fisk}, L.~A., {et~al.} 2000, \jgr, 105,
  27217, \dodoi{10.1029/1999JA000358}

\bibitem[{{Wilhelm} {et~al.}(1998){Wilhelm}, {Marsch}, {Dwivedi}, {Hassler},
  {Lemaire}, {Gabriel}, \& {Huber}}]{Wilhelm_1998ApJ}
{Wilhelm}, K., {Marsch}, E., {Dwivedi}, B.~N., {et~al.} 1998, \apj, 500, 1023,
  \dodoi{10.1086/305756}

\bibitem[{{Withbroe}(1988)}]{Withbroe_1988ApJ}
{Withbroe}, G.~L. 1988, \apj, 325, 442, \dodoi{10.1086/166015}

\bibitem[{{Withbroe} \& {Noyes}(1977)}]{Withbroe_1977ARA&A}
{Withbroe}, G.~L., \& {Noyes}, R.~W. 1977, \araa, 15, 363,
  \dodoi{10.1146/annurev.aa.15.090177.002051}

\bibitem[{{Wood} {et~al.}(2005){Wood}, {M{\"u}ller}, {Zank}, {Linsky}, \&
  {Redfield}}]{Wood_2005ApJ}
{Wood}, B.~E., {M{\"u}ller}, H.~R., {Zank}, G.~P., {Linsky}, J.~L., \&
  {Redfield}, S. 2005, \apjl, 628, L143, \dodoi{10.1086/432716}

\bibitem[{{Wood} {et~al.}(2021){Wood}, {M{\"u}ller}, {Redfield}, {Konow},
  {Vannier}, {Linsky}, {Youngblood}, {Vidotto}, {Jardine},
  {Alvarado-G{\'o}mez}, \& {Drake}}]{Wood_2021ApJ}
{Wood}, B.~E., {M{\"u}ller}, H.-R., {Redfield}, S., {et~al.} 2021, \apj, 915,
  37, \dodoi{10.3847/1538-4357/abfda5}

\bibitem[{{Yoshida} {et~al.}(2019){Yoshida}, {Suematsu}, {Ishikawa}, {Okamoto},
  {Kubo}, {Kano}, {Narukage}, {Bando}, {Winebarger}, {Kobayashi}, {Trujillo
  Bueno}, \& {Auch{\`e}re}}]{Yoshida_2019ApJ}
{Yoshida}, M., {Suematsu}, Y., {Ishikawa}, R., {et~al.} 2019, \apj, 887, 2,
  \dodoi{10.3847/1538-4357/ab4ce7}

\bibitem[{{Zangrilli} {et~al.}(2002){Zangrilli}, {Poletto}, {Nicolosi}, {Noci},
  \& {Romoli}}]{Zangrilli_2002ApJ}
{Zangrilli}, L., {Poletto}, G., {Nicolosi}, P., {Noci}, G., \& {Romoli}, M.
  2002, \apj, 574, 477, \dodoi{10.1086/340942}

\bibitem[{{Zank} {et~al.}(2021){Zank}, {Zhao}, {Adhikari}, {Telloni}, {Kasper},
  \& {Bale}}]{Zank_2021PhPl}
{Zank}, G.~P., {Zhao}, L.~L., {Adhikari}, L., {et~al.} 2021, Physics of
  Plasmas, 28, 080501, \dodoi{10.1063/5.0055692}

\end{thebibliography}

\end{document}